\documentclass[letterpaper,twocolumn,10pt]{article}
\usepackage{usenix-2020-09}

\usepackage{authblk}
\usepackage{tikz}
\usepackage{amsmath}
\usepackage{amssymb}
\usepackage{pifont}
\usepackage{filecontents}

\usepackage{algorithm}
\usepackage{algpseudocode}

\usepackage{caption}
\usepackage{graphicx}
\usepackage{diagbox}
\usepackage{float} 
\usepackage{subcaption}
\usepackage{threeparttable}

\newtheorem{thm}{Theorem}

\newtheorem{assumption}{Assumption}
\newtheorem{definition}{Definition}

\newenvironment{thmprime}[1]
  {\renewcommand{\thethm}{\ref{#1}}%
   \addtocounter{thm}{-1}%
   \begin{thm}}
  {\end{thm}}
\newenvironment{thmprime2}[1]
  {\renewcommand{\thethm}{\ref{#1}}%
   \addtocounter{thm}{-1}%
   \begin{thm}}
  {\end{thm}}

\begin{document}

\date{}

\title{\Large \bf Towards Understanding and Enhancing Security of Proof-of-Training\\
   for DNN Model Ownership Verification}

\author[$\dag$]{Yijia Chang}
\author[$\dag$]{Hanrui Jiang}
\author[$\,\,\,\,\,$]{Chao Lin\thanks{Corresponding author: Chao Lin.}}
\author[$\ddag$]{Xinyi Huang}
\author[$\ddag$]{Jian Weng}
\affil[$\dag$]{The Hong Kong University of Science and Technology (Guangzhou)}
\affil[*]{Fujian Normal University$\qquad\qquad\qquad$\textsuperscript{\ddag}{Jinan University}}
\affil[ ]{\textit{\{ychang847, hjiang260\}@connect.hkust-gz.edu.cn, linchao91@fjnu.edu.cn, \{xyhuang81,cryptjweng\}@gmail.com}}




\maketitle

\begin{abstract}
The great economic values of deep neural networks (DNNs) urge AI enterprises to protect their intellectual property (IP) for these models. Recently, proof-of-training (PoT) has been proposed as a promising solution to DNN IP protection, through which AI enterprises can utilize the record of DNN training process as their ownership proof. To prevent attackers from forging ownership proof, a secure PoT scheme should be able to distinguish honest training records from those forged by attackers. Although existing PoT schemes provide various distinction criteria, these criteria are based on intuitions or observations. The effectiveness of these criteria lacks clear and comprehensive analysis, resulting in existing schemes initially deemed secure being swiftly compromised by simple ideas. 
In this paper, we make the first move to identify distinction criteria in the style of \textit{formal methods}, so that their effectiveness can be explicitly demonstrated.
Specifically, we conduct systematic modeling to cover a wide range of attacks and then theoretically analyze the distinctions between honest and forged training records.
The analysis results not only induce a universal distinction criterion, but also provide detailed reasoning to demonstrate its effectiveness in defending against attacks covered by our model. 
Guided by the criterion, we propose a generic PoT construction that can be instantiated into concrete schemes. This construction sheds light on the realization that trajectory matching algorithms, previously employed in data distillation, possess significant advantages in PoT construction. Experimental results demonstrate that our scheme can resist attacks that have compromised existing PoT schemes, which corroborates its superiority in security.

\end{abstract}

\section{Introduction}

Driven by the ever-increasing economic value of deep neural networks (DNNs), how to protect the intellectual property (IP) of DNN models has been a great concern for artificial intelligence (AI) enterprises~\cite{oliynykKnowWhatYou2023}.
This concern becomes particularly strong since various model stealing (extraction) attack methods are proposed~\cite{Wang2018StealingHI,chenDDAEDefensePenetratingModel2023,orekondyKnockoffNetsStealing2019}.
To address this concern, a common solution is to devise DNN model ownership verification schemes, by which a verifier (e.g., patent office or court) can distinguish the legal model owner from attackers. These schemes are expected to be \textit{secure} so that attackers can not cheat the verifier to obtain model ownership.

Recently, proof-of-training (PoT)~\cite{jiaProofofLearningDefinitionsPractice2021a,gargExperimentingZeroKnowledgeProofs2023,sunZkDLEfficientZeroKnowledge2023,liuProvenanceTrainingTraining2023a} has been proposed as a promising solution for model ownership verification, also known as proof-of-learning or provenance-of-training.
Differing from other schemes that verify the \textit{features} of the target model (e.g., watermarking~\cite{chenDeepMarksSecureFingerprinting2019,darvishrouhaniDeepSignsEndtoEndWatermarking2019,uchidaEmbeddingWatermarksDeep2017,wangFreeFinetuningPlugandPlay2023,ganRobustModelWatermark2023} and fingerprinting~\cite{cao2021a,lukas2021a,wang2021a,zhao2020a}), PoT schemes examine the knowledge of the training record to distinguish the model owner from attackers.
Typically, a training record consists of training data, training algorithms, and a training trajectory from the initial model state to the final model state. If a party has a valid training record for a particular model, then it is recognized as the model owner. Note that attackers can steal the model but does not perform its training process. Hence, compared with manipulating model features, attackers face more challenges to forge their training records and may even need to pay much more effort than honest training~\cite{jiaProofofLearningDefinitionsPractice2021a}. This greatly reduces the profits of attackers from stealing the model. Therefore, in this paper, we focus on PoT schemes and consider their design for DNN model ownership verification.




To ascertain the authenticity of training records as either honest or forged, researchers have proposed multiple PoT schemes. These schemes employ various criteria to examine the distinctions between honest and forged training records. The different types of examined distinctions divide existing PoT schemes into two categories. One is \textit{retraining-based} category~\cite{jiaProofofLearningDefinitionsPractice2021a,sunZkDLEfficientZeroKnowledge2023,gargExperimentingZeroKnowledgeProofs2023} that examines whether the training trajectory can be generated by retraining from training data and algorithms. The other is \textit{statistics-based} category. 
Since training data may be private, this category does not examine training data but only assesses whether the statistical metrics of the training trajectory satisfy pre-specified requirements~\cite{liuProvenanceTrainingTraining2023a}.
These works provide useful ideas to verify the validity of training records, which greatly advance the development of PoT.




However, the security of existing PoT schemes heavily relies on intuitive conjectures and experimental observations. We are neither sure about what kinds of attacks can be resisted by existing schemes nor about the potential security vulnerabilities of PoT. As a result, existing schemes that are initially thought to be secure may end up being compromised by more tricky attacks. For example, researchers have breached the security of retraining-based PoT schemes by forging a training record that supports retraining~\cite{zhangAdversarialExamplesProofofLearning2022,fangProofofLearningCurrentlyMore2023}, and this vulnerability has not been fixed yet. As for the statistics-based category, we show a successful attack against it in Section \ref{sec:modeling}. These ever-emerging forging methods will diminish people's confidence in the security of PoT schemes.

Faced with this current situation, we believe that \textit{formal methods} are essential for further enhancement of PoT security. Formal methods are techniques used to model a complex system as logical entities, so that we can examine the security of this system in a more thorough fashion than piecemeal enumeration~\cite{formal}.
Based on the system modeling, we can obtain various useful conclusions through theoretical analysis, e.g., ruling out a range of attacks from success.  
Even though the model must have some assumptions, the well-defined assumptions can provide a clue for further improvement.  

In this paper, we are motivated to follow the spirit of formal methods and apply them to the study of PoT security.
Although formal methods have been used in many areas, to the best of our knowledge, this is the first work to consider the application of formal methods in PoT security.
By modeling the PoT mechanism, we aim to answer two questions through theoretical analysis: 1) \textit{What are the key distinctions between honest and forged training records?} and 2) \textit{How to detect these distinctions to prevent attackers from forging training records?}
The answer to these questions can not only provide a comprehensive understanding of PoT security but also reveal a clear guideline to enhance the security of PoT schemes.
For example, our modeling of attacks helps to pinpoint the focus of defense within a limited but reasonable range of attacks, and our analysis of forged training record's features can enlighten more methods to detect them.
Below we summarize the main results and key contributions of this paper.
\begin{itemize}
    \item[$\diamond$] \textbf{Comprehensive modeling of attacks against PoT.} By formalizing the training record and threat model, we characterize both the capabilities and incapabilities of attackers. Such a characterization enables us to identify ``non-trivial'' attacks that are neither weak enough to be detected by existing PoT schemes, nor too strong to be feasible in practice. Among these non-trivial attacks, we discover that they share common features and can be classified into two types, namely algorithm manipulation and data fabrication attacks. By characterizing the features of these two attack types, we can cover potentially infinite attacks in our model. Particularly, we note that the two types can cover successful attacks against existing retraining-based and statistics-based PoT schemes.
    \item[$\diamond$] \textbf{A universal criterion for detecting attacks.} We adopt information theory to analyze the key distinctions between honest training records and training records forged by the aforementioned non-trivial attacks. The analysis results show that training records forged by different attacks can be detected by a universal criterion. In particular, when compared with honest training records, forged training algorithms and trajectory can only induce \textit{low-fidelity} data that either fail to satisfy retraining requirements or deviate significantly from the target distribution. This universal criterion can help to enhance the security of PoT schemes.
    \item[$\diamond$] \textbf{A generic PoT construction without using training data.} Guided by the above criterion, we propose a generic PoT construction based on trajectory matching algorithms~\cite{cazenavetteDatasetDistillationMatching2022,cuiScalingDatasetDistillation2023}. These algorithms enable the verifier to synthesize data that induce a similar trajectory with the same training algorithms. Instead of examining training data, the verifier can detect attacks by checking the fidelity of synthetic data. This construction establishes a connection between trajectory matching algorithms and PoT, which represents an intersection of independent interests within both research lines. Notably, these algorithms were previously applied in the context of data distillation, and this paper sheds light on their application in model ownership verification. 
    \item[$\diamond$] \textbf{Experimental security evaluations against various attacks.} To demonstrate the security of our construction, we instantiated a concrete PoT scheme using the trajectory matching algorithm with memory efficiency~\cite{cuiScalingDatasetDistillation2023} and implemented it on classical DNN models and datasets. Then we simulated various forging methods, including those successful attacks against existing PoT schemes. The experimental results show that our scheme effectively distinguishes the model owner from attackers under these attacks, which corroborates its superiority in terms of its security. In addition, the defense effects become even stronger with the growth of model size and dataset complexity. Since the commercial scenario mainly involves large DNN models trained over complex datasets, our PoT scheme has the potential to be applied in the real world.
\end{itemize}

The rest of this paper is organized as follows. Section \ref{sec:overview} provides a technical overview of this paper. Section \ref{sec:modeling} presents our modeling of attack methods, and Section \ref{sec:criterion} analyzes the criterion for detecting these attacks. Guided by the criterion, Section \ref{sec:scheme} introduces our PoT construction that can be instantiated into concrete PoT schemes. Next, Section \ref{sec:experiment} evaluates the security of the PoT scheme. At last, Section \ref{sec:related} reviews related work and Section \ref{sec:conclusion} concludes this paper.

\section{Technical Overview}
\label{sec:overview}
\begin{figure}[htbp]
	\centering
	\includegraphics[width=\linewidth]{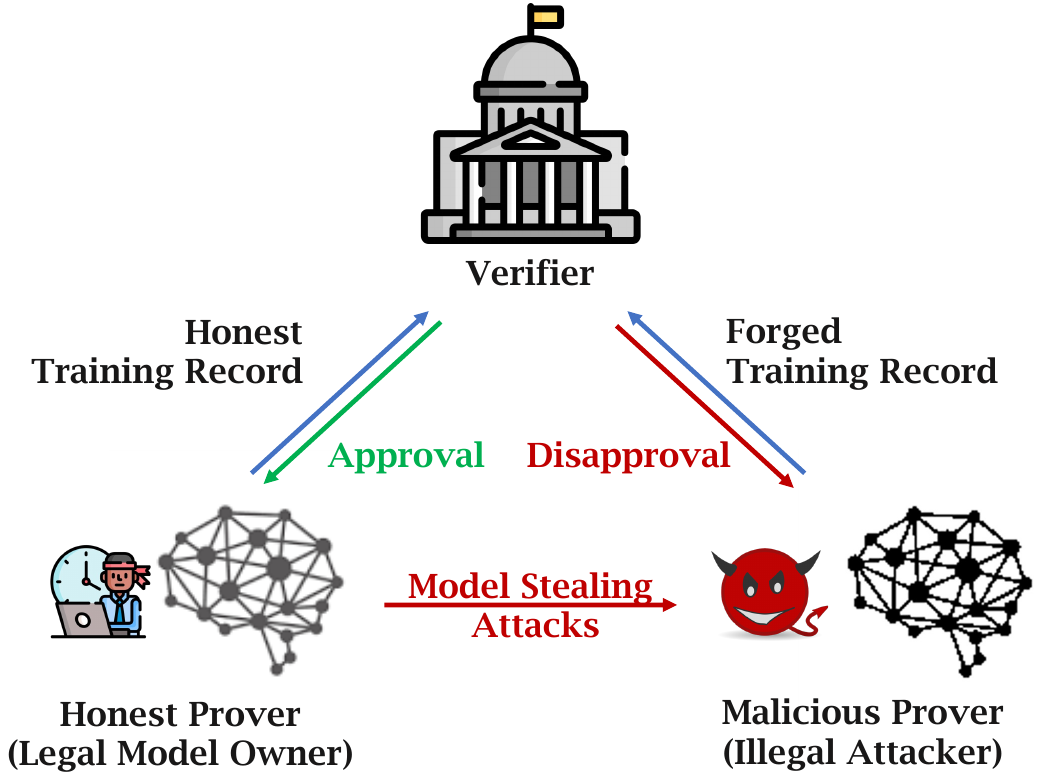}
	\caption{PoT-based model ownership verification.}
	\label{fig:framework}
\end{figure}

When a party claims its ownership of a DNN model, PoT aims to examine the ownership by verifying whether the party is the real trainer of this DNN model. As illustrated in Figure \ref{fig:framework}, this verification involves three types of stakeholders: \textit{verifier} is an honest official institution (e.g., patent office or court) who is responsible for examining model ownership; \textit{honest prover} performs the training of a DNN model and generates a record of the training process; and \textit{malicious prover} wishes to illegally claim model ownership after stealing a model from the honest prover.

To verify model ownership, the idea behind PoT is that the honest prover does have the record of the training process, while malicious provers do not. Typically, a training record consists of training algorithms as well as their inputs (i.e., training data) and outputs (i.e., a training trajectory from the initial model to the final model). Based on this idea, a secure PoT scheme should distinguish an honest prover from malicious provers according to their training records.
By the above description, the central questions in designing a secure PoT scheme include: 1) \textit{What are the key distinctions between honest and malicious training records?} and \textit{2) How to detect these distinctions?}

For the first question, the two PoT categories mentioned above~\cite{jiaProofofLearningDefinitionsPractice2021a,gargExperimentingZeroKnowledgeProofs2023,sunZkDLEfficientZeroKnowledge2023,liuProvenanceTrainingTraining2023a} explore the answers from different perspectives. In the retraining-based PoT schemes~\cite{jiaProofofLearningDefinitionsPractice2021a,gargExperimentingZeroKnowledgeProofs2023,sunZkDLEfficientZeroKnowledge2023}, the training trajectory can be retrained from the training algorithms over training data in an honest record but cannot do so in a malicious record. In the statistics-based PoT schemes~\cite{liuProvenanceTrainingTraining2023a}, through observation and experimental verification, several statistical characteristics are identified to distinguish between honest and malicious training trajectories.


Although existing PoT schemes are shown to be effective against some particular attacks, they may be insecure against more tricky attacks. For the retraining-based category, existing works have found successful attacks against it~\cite{fangProofofLearningCurrentlyMore2023,zhangAdversarialExamplesProofofLearning2022}. The statistics-based category, proposed recently, remains unbreached yet. Nonetheless, by slightly modifying a known attack, we successfully attack the statistics-based category in a more realistic setting. The details of this attack and its effect are presented in Section \ref{sec:successfulattack}.


Building on the above investigations, we apply formal methods to deepen our understanding of attack methods and guide secure PoT scheme design. Specifically, to answer the first question, we take the following three steps.

\begin{itemize}
    \item First, we formalize the training record and the threat model of malicious provers, with a focus on the goal, capabilities, and incapabilities of malicious provers. Particularly, these incapabilities play the role of security assumption in our modeling.
    \item Second, we model those ``non-trivial'' attack methods that are feasible under the capabilities of malicious provers but can not be detected by existing PoT schemes.
    \item Third, we analyze the key distinctions between honest training records and malicious training records forged by those non-trivial attacks.
\end{itemize}


Through these three steps, we provide a thorough classification of potential attacks and obtain some useful analysis results to answer the first question. Particularly, attacks are divided into three categories: 1) weak attacks that can be detected by existing PoT schemes, 2) strong attacks that are out of the range of attackers' capabilities, and 3) non-trivial attacks. Particularly, we put the focus on \textit{non-trivial} attacks and analyze the key distinctions between honest and forged training records under each type of non-trivial attacks.
One type is algorithm manipulation attacks, under which we find that the training trajectory and training data from the \textit{honest} training record are more dependent on each other. More formally, we use mutual information $I$ to measure the dependence and prove the following theorem. 
\begin{thm}[Informal]
\label{theo:dis}
    Given two datasets $D$ and $D^{(M)}$ sampled from the same distribution, suppose that $\mathbb{T}_T$ is trajectory output by honest training algorithms and $\mathbb{T}_T^{(M)}$ is trajectory output by forged training algorithms, then we have 
    \begin{equation}
        I(D;\mathbb{T}_T)> I(D^{(M)};\mathbb{T}_{T}^{(M)}).
    \end{equation}
\end{thm}
\noindent The other type is data fabrication attacks, under which the key distinction between training records is the data distribution. The forged training data have a different distribution from the honest training data. With these analysis results, we can design a secure PoT scheme by checking these distinctions.

Nonetheless, even though we have the above guidelines, answering the second question (i.e., detection of these distinctions) still faces two challenges.
The first challenge is data privacy concern. Although the verifier is assumed to be honest, the laws and regulations about data privacy may forbid the provers to upload their sensitive training data to the verifier.
The second challenge is high-dimensional trajectory. For the algorithm manipulation attacks, the verifier needs to measure the mutual information between the data and trajectory. Although there are several mutual information estimation techniques~\cite{MIME}, the dimension of a trajectory is too high to make an accurate measurement, especially for large DNN models.

To tackle the above challenges, we show that there exists a universal criterion for detecting the above attacks, without using the training data. Through detailed reasoning in Section \ref{sec:criterion}, we show that if the training algorithms and trajectory come from malicious provers, it is hard to recover \textit{high-fidelity} training data from them such that 1) operating training algorithms over training data leads to a similar trajectory and 2) the training data comes from target data distribution. Guided by such a criterion, the verifier can detect the above attacks in two steps: 1) generating synthetic data so that training over them produces a similar trajectory and 2) evaluating whether the synthetic data align with the target data distribution. Notably, these two steps do not rely on the training data and thus can help to address the data privacy concern.

\begin{figure}[htbp]
	\centering
	\includegraphics[width=\linewidth]{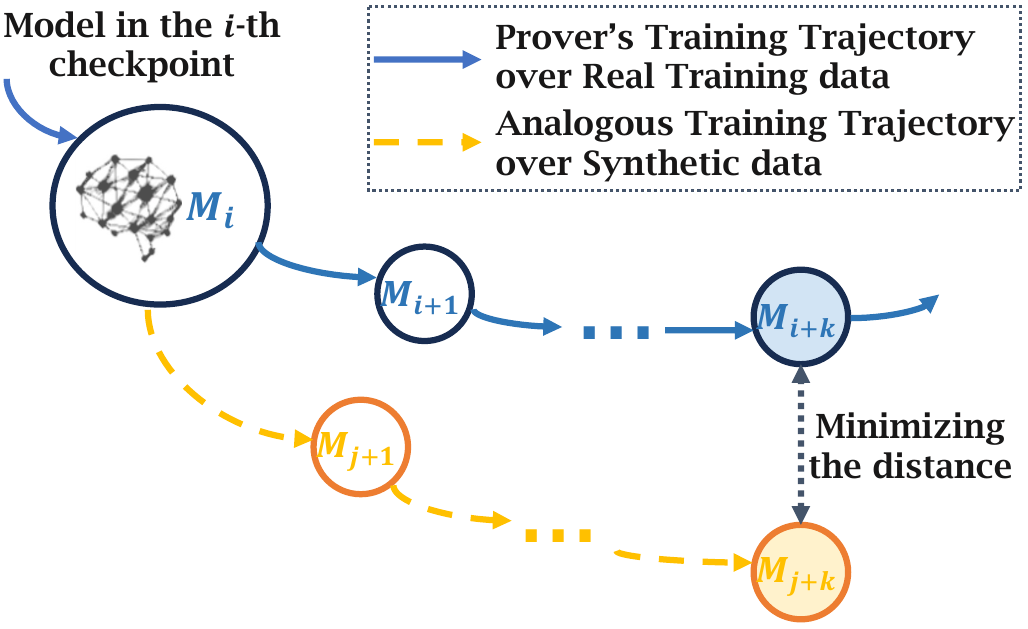}
	\caption{Trajectory matching algorithm in the scenario of PoT construction. The solid blue lines represent the training trajectory from the prover's training record, which should be trained over training data. After sampling a fragment from this trajectory, say from $M_i$ to $M_{i+k}$, the verifier trains an analogous training trajectory over synthetic data starting from $M_i$, which is depicted with dashed yellow lines. By minimizing the distance between the destinations of these two trajectories, the synthetic data will perform similar behaviors for DNN training, i.e., training over synthetic data produces a trajectory similar to the prover's trajectory.}
	\label{fig:trajectory}
\end{figure}

Based on the above ideas, we propose a generic PoT construction. To conduct the first step, we adopt \textit{trajectory matching} algorithms to recover the synthetic data $S$ from the training algorithms and trajectory. As illustrated in Figure \ref{fig:trajectory}, the trajectory matching algorithms are designed to minimize the distance between two trajectories trained over different data.
To apply trajectory matching algorithms in our scenario, one trajectory is set to be the trajectory from provers, while the other trajectory is trained over synthetic data by the verifier. In this way, the synthetic data can be used to generate a similar trajectory with training algorithms. As for the second step, we assume that the verifier has a small portion of test data sampled from the target data distribution. To evaluate whether the synthetic data comes from the target distribution, the verifier trains several new DNN models over synthetic data $S$ and observes the average accuracy of these new models on test data. Among multiple proofs from different provers, the proof leading to the highest average accuracy is recognized as the honest prover.

To validate the security of our PoT scheme, we implement a prototype and corroborate its security through extensive experiments upon various DNN models and datasets. The experiment results demonstrate that our scheme can defend against various attacks, including those successful attacks against existing PoT schemes.

\section{Modeling of Attacks against PoT}
\label{sec:modeling}

In this section, we provide a comprehensive modeling of attacks against PoT. To this end, we formalize the training record in Section \ref{sec:modeling-record}, model the malicious provers in Section \ref{sec:modeling-threat}, and characterize the methods of forging training records in Section \ref{sec:modeling-attack}.


\subsection{Formalization of Training Record}
\label{sec:modeling-record}
To formalize the record of training process, we review the training process and study those contents that can be saved as records. On a high level, DNN training usually starts from an initial model and produces a final model via gradient descent algorithms. The algorithm for model initialization, denoted by $\mathbb{I}_A$, is either to sample a random value from a particular distribution or to choose an existing model that has shown good accuracy. After the model is initialized, DNN training will update models in a number of epochs following the gradient descent algorithm. During the multiple epochs of training process, it is common practice to save model states at some checkpoints. We call the sequence of these saved model states \textit{training trajectory}, which is formally defined as follows.

\begin{definition}
\label{def:trajectory}
    \textbf{Training trajectory} $\mathbb{T}_T$ is a sequence of $n$ model states $\{M_1,\ldots,M_n\}$, where $M_1$ is the initial model and $M_i$ is the saved model state in the $i$-th checkpoint for $2\leq i\leq n$. Particularly, the final model state $M_n$ is called the \textbf{tail} of training trajectory, and the value of $n$ is called the \textbf{length} of training trajectory.
\end{definition}


To update models from one checkpoint to the next checkpoint, DNN training needs to compute a model update by learning from training data $D_i$ that are sampled from the target distribution $\mathcal{D}$. We call the concrete method of computing model updates \textit{training algorithm} and define it as follows.
\begin{definition}
\label{def:algorithm}
    \textbf{Training algorithm} $\mathbb{T}_{A,i}$ for checkpoint $i$ is a function that maps model state $M_{i}$ and data $D_i$ to model update $\Delta_i$.
\end{definition}
In practice, a training algorithm can exist in the form of an executable program that takes $M_i,D_i$ as input and outputs $\Delta_i$. 
Notably, this program involves many details, including model architecture, optimizer, loss function, hyper-parameters, etc. These details, regarded as the parameters of training algorithms, are not explicitly written in symbols.

Under ideal circumstances, we should have $M_{i+1}=M_i+\Delta_i$.
Nonetheless, we note that there usually exists a gap between the empirical results of model updates (i.e., $M_{i+1}-M_i$) and the theoretical outputs of $\mathbb{T}_{A,i}$, which is caused by a random noise from accelerator devices such as graphics processing units~\cite{jiaProofofLearningDefinitionsPractice2021a}. While such a noise usually has a bounded value, it still affects the training trajectory.
More formally, suppose that the random noise for the $i$-th checkpoint is $z_i$ with a $B$-bounded value, then we have
\begin{equation}
\label{eq:honest-training}
    M_{i+1}=M_{i}+\Delta_i+z_i=M_{i}+\mathbb{T}_{A,i}(M_{i},D_i)+z_i.
\end{equation}

Based on the above discussions, the training record is simply the combination of these ingredients for DNN training.

\begin{definition}
\label{def:record}
    \textbf{Training record} $\,\mathbb{T}_{R,n}$ with length $n$ is a quadruple $(\mathbb{I}_A,\mathbb{T}_A,\mathbb{T}_T,D)$ with the following properties:
    \begin{enumerate}
        \item $\mathbb{I}_A$ is an initialization algorithm to generate $M_1$.
        \item $\mathbb{T}_A=\{\mathbb{T}_{A,1},\ldots,\mathbb{T}_{A,n-1}\}$ is a sequence of $n-1$ training algorithms.
        \item $\mathbb{T}_T=\{M_1,\ldots,M_n\}$ is a length-$n$ training trajectory.
        \item $D=\{D_1,\ldots,D_{n-1}\}$ is a sequence of $n-1$ datasets.
        \item For $1\leq i\leq n-1$, $M_{i+1}=M_i+\mathbb{T}_{A,i}(M_i,D_i)+z_i$, where $z_i$ is a noise with $B$-bounded value.
    \end{enumerate}
\end{definition}

\subsection{Modeling of Malicious Provers}
\label{sec:modeling-threat}

\subsubsection{Attack Goal}
The goal of malicious provers is to forge a training record on some target model so that the verifier can not distinguish the honest training record from the forged training record. 
Formally, let $\mathbb{T}_{R,n}^{(H)}$ (resp. $\mathbb{T}_{R,m}^{(M)}$) denote the training record from the honest (resp. malicious) prover. The malicious prover succeeds if the verifier approves $\mathbb{T}_{R,m}^{(M)}$ rather than $\mathbb{T}_{R,n}^{(H)}$.

However, if the malicious prover is willing to invest unlimited efforts in the attack, there exist trivial attack methods that are guaranteed to succeed. The malicious prover simply needs to search through all possible training records and select those with the tail being the target model. After enough trials, these training records must include the honest training record.
Still, in this case, the successful attacks become somewhat meaningless: it would be more valuable to honestly train the model rather than forging the training record with much more effort.

In view of the above, we follow existing PoT scheme~\cite{jiaProofofLearningDefinitionsPractice2021a} and consider the malicious prover who aims to forge the training record with \textit{less} effort than honest provers. To compare the efforts of provers, we measure the ``cost'' paid by the provers to generate training records. Since the training record (in particular, the training trajectory) is usually the output of some algorithm(s), we define the cost over these algorithm(s). More specifically, we use a positive real number to represent how much these algorithms cost. In practice, this number can be the running time or the volume of computation resources required to execute these algorithms. Note that the cost of an algorithm usually varies across distinct inputs, especially when the sizes of these inputs are different. Hence, the definition of cost should also take the inputs into account. Based on the above considerations, we define the cost as follows.

\begin{definition}
    The \textbf{cost} is a function $Cost(x_1,x_2)$ that maps algorithms $x_1$ and their inputs $x_2$ to a positive real number that represents the ``effort'' of performing $x_1$ over $x_2$. When the inputs $x_2$ are fixed and clear from the context, we will omit the $x_2$ and use the notation $Cost(x_1)$ instead.
\end{definition}

Throughout this paper, we use the running time on a fixed machine to evaluate and compare the costs of honest and malicious provers. Next, we take the cost of honest prover as an example to illustrate this notion and defer the cost evaluation of malicious provers to the modeling of attack methods in Section \ref{sec:modeling-attack}. Given $\mathbb{T}_{R,n}^{(H)}=(\mathbb{I}_A^{(H)},\mathbb{T}_T^{(H)},\mathbb{T}_A^{(H)},D^{(H)})$, the honest prover runs the initialization algorithm $\mathbb{I}_A^{(H)}$ and the training algorithms $\mathbb{T}_A^{(H)}$ to output this training record.
Let $\mathsf{input}_1$ and $\mathsf{input}_2$ denote the inputs of $\mathbb{I}_A^{(H)}$ and $\mathbb{T}_A^{(H)}$, respectively.
Then the effort of honest prover is $Cost(\mathbb{I}_A^{(H)},\mathsf{input}_1)+Cost(\mathbb{T}_A^{(H)},\mathsf{input}_2)$. More concretely, suppose that the honest prover takes $E$ epochs for training and each epoch takes running time $t$ on average, then we have $Cost(\mathbb{T}_A^{(H)},\mathsf{input}_2)=E\cdot t$. Notably, for a specific honest prover, its inputs $\mathsf{input}_1$ and $\mathsf{input}_2$ that are used to generate $\mathbb{T}_{R,n}^{(H)}$ are fixed. For example, the $\mathsf{input}_2$ usually consists of the training datasets in $D^{(H)}$ and the first $n-1$ models in $\mathbb{T}_T^{(H)}$. Hence, we simply use $Cost(\mathbb{I}_A^{(H)})+Cost(\mathbb{T}_A^{(H)})$ to denote the cost of the honest prover in the subsequent contents.

\subsubsection{Attack Capability}
In order to achieve the above goal, we assume malicious provers have the following adversarial capabilities.
\begin{itemize}
    \item Malicious provers have access to the target model;
    \item Malicious provers have access to the training datasets used by honest provers and can even sample new datasets from the same distribution;
    \item Malicious provers have the full knowledge of the initialization and training algorithms, including all details such as initialization strategy, model architecture, optimizer, loss function, and even the seed for randomization. 
\end{itemize}

We remark that the full access to training datasets is one key difference between our setting and existing statistics-based PoT schemes~\cite{liuProvenanceTrainingTraining2023a}. They assume that the size of datasets owned by malicious provers is much smaller than that of honest provers (e.g., no greater than $10\%$). This assumption may be inconsistent with the real world, e.g., honest provers could adopt some public training datasets that can be accessed by malicious provers.

\subsubsection{Attack Incapability}
To exclude strong attacks that may be infeasible in practice, we make two assumptions to restrict the capability of malicious provers.

First, we assume that the malicious prover cannot find a sequence of training algorithms that require less effort than that of honest prover, if it can update the initial model to the target model \textit{without utilizing the target model}. The reasons are two-fold. On one hand, if the malicious prover can find such an algorithm sequence, it can honestly train its own model by simply running this algorithm sequence, rather than stealing the target model and forging the training record.
On the other hand, if such an algorithm sequence exists, the honest prover is usually smart enough to use the state-of-the-art training algorithms.
To formulate this assumption, we need to formally define ``utilizing the target model''. When we say an algorithm sequence utilizes the target model $M$, we imply that $M$ is an intrinsic (hyper-)parameter of the algorithms and will lead the algorithm output (i.e., the updated model) towards $M$. In other words, the output will eventually be $M$ due to the existence of $M$ in the algorithms, no matter what the inputs are. Let $\mathbb{T}_{R,m}^{(M)}=(\mathbb{I}_A^{(M)},\mathbb{T}_T^{(M)},\mathbb{T}_A^{(M)},D^{(M)})$ denote the length-$m$ training record generated by the malicious prover. Then ``$\mathbb{T}_A^{(M)}$ utilizes the target model $M$'' is defined as: for any $D^{(M)}$, there exists an $m$ that is large enough so that the distance between trajectory's tail and the target model $d(M_m,M)$ is less than a small threshold $\epsilon$.
Based on the above discussions, we formulate this assumption as follows. 
\begin{assumption}
    \label{assu:strong}
    Given a target model $M$ generated by the sequence of honest training algorithms $\mathbb{T}_A^{(H)}$, if a sequence of training algorithms $\mathbb{T}_A^{(M)}$ can update an honest initial model $M_1$ to $M$ using datasets $D^{(M)}$, then either $Cost(\mathbb{T}_A^{(M)},\{\mathbb{T}_T^{(M)},D^{(M)}\})>Cost(\mathbb{T}_A^{(H)})$ or $d(M_m,M)<\epsilon$ holds for any $D^{(M)}$ and threshold $\epsilon$ when $m$ is large enough.
\end{assumption}

Second, we assume that the malicious prover cannot find a dataset $D_i$ to guarantee that $\mathbb{T}_{A,i}^{(M)}(M_i,D_i)$ outputs a given value (say $\Delta$) with less cost than running $\mathbb{T}_{A,i}^{(M)}$, unless $\mathbb{T}_{A,i}^{(M)}$ outputs $\Delta$ for any $D_i$. In other words, when the output of $\mathbb{T}_{A,i}^{(M)}$ is not a constant value, if there exists an algorithm $f$ who can find $D_i$ on input $\mathsf{input}_f$ so that $\mathbb{T}_{A,i}^{(M)}(M_i,D_i)=\Delta$, then $Cost(f,\mathsf{input}_f)>Cost(\mathbb{T}_{A,i}^{(M)},\{M_i,D_i\})$. The reason is that verifying the answer to a problem usually costs less than solving this problem. Hence, to find $D_i$ such that $\mathbb{T}_{A,i}^{(M)}(M_i,D_i)=\Delta$, the algorithm $f$ needs to costs higher than just running $\mathbb{T}_{A,i}^{(M)}$ once. In fact, existing works have shown that the cost of algorithm $f$ is generally much higher than $\mathbb{T}_{A,i}^{(M)}$~\cite{jiaProofofLearningDefinitionsPractice2021a}. We summarize this assumption as follows.

\begin{assumption}
    \label{assu:strong-2}
    Given a value $\Delta$, if the output of $\mathbb{T}_{A,i}^{(M)}$ varies with respect to different $D_i$s, then an algorithm $f$ that can find $D_i$ on input $\mathsf{input}_f$ so that $\mathbb{T}_{A,i}^{(M)}(M_i,D_i)=\Delta$ must have $Cost(f,\mathsf{input}_f)>Cost(\mathbb{T}_{A,i}^{(M)},\{M_i,D_i\})$.
\end{assumption}

\subsection{Modeling of Attack Methods}
\label{sec:modeling-attack}

\subsubsection{Taxonomy of Potential Attacks}


Next, we model the attack methods for forging training records. Our modeling aims to focus on those non-trivial attack methods that 1) can not be easily detected by existing defense methods and 2) are feasible within the capabilities of attackers.
To this end, we first exclude two types of ``weak'' attacks in the subsequent modeling of non-trivial attacks.

The first type of weak attacks outputs a forged training record with \textit{incorrect structure}. The correct structure of a length-$n$ training record should consist of a length-$n$ trajectory, an initialization algorithm, a sequence of $n-1$ training algorithms and/or datasets. Each training algorithm should be in the form of an executable program. Its input is the preceding model and a training dataset, and its output is the model update. The violation of these requirements will be marked as ``incorrect structure'' and excluded from verification.

The second type of weak attacks forges a \textit{dishonest initialized model} $M_1$. To detect this type of attacks, we restrict the honestly initialized model to two possible cases. In one case, the initialization algorithm samples $M_1$ from a random distribution and a forged $M_1$ can be detected by statistical tests. In the other case, the initialization algorithm adopts an existing well-trained model as the initialized model. To detect a forged $M_1$ in this case, a common practice is to check whether this well-trained model has an ownership proof~\cite{jiaProofofLearningDefinitionsPractice2021a}. If the initialization algorithm does not fall into these two cases or the generation of $M_1$ does not honestly follow the initialization algorithm, we identify a training record as ``dishonest initialization''.

In the subsequent modeling of attack methods, we exclude these two attack methods and focus on forged training records with correct structure and honest initialization.
We first note that the potential infinite attack methods can be classified from two orthogonal perspectives. 

\begin{figure*}[htbp]
	\centering
	\begin{subfigure}{0.23\linewidth}
		\centering
		\includegraphics[width=0.95\linewidth]{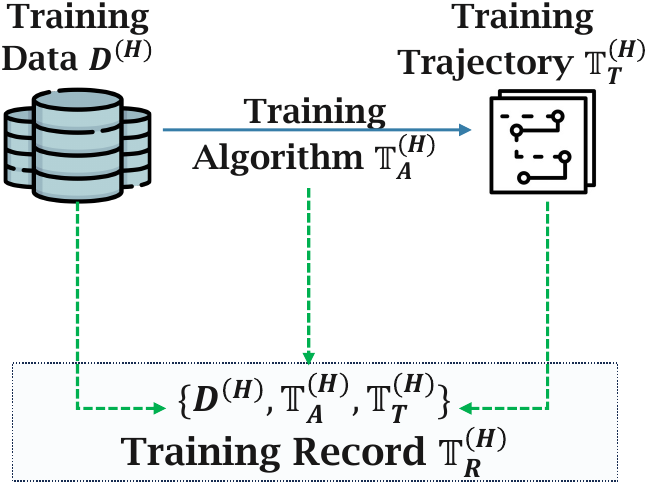}
		\caption{Honest DNN Training}
		\label{fig:model-training}
	\end{subfigure}
	\centering
	\begin{subfigure}{0.29\linewidth}
		\centering
		\includegraphics[width=0.95\linewidth]{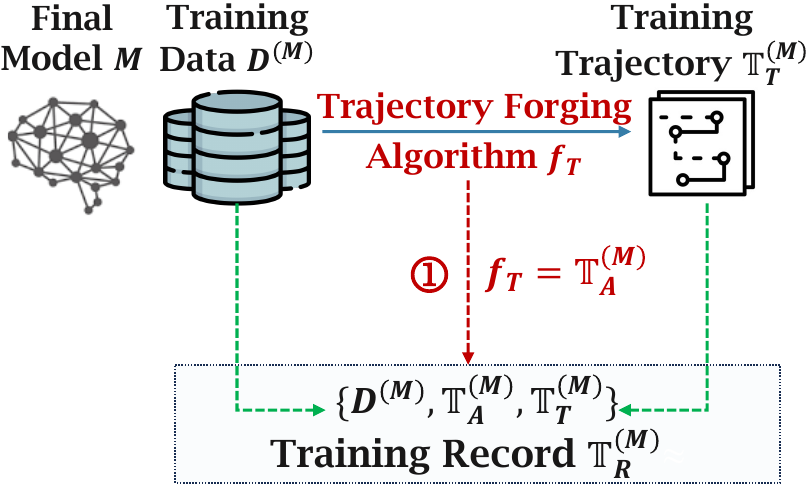}
		\caption{Forward-Direction Attacks}
		\label{fig:model-algorithm}
	\end{subfigure}
	\centering
	\begin{subfigure}{0.44\linewidth}
		\centering
		\includegraphics[width=0.95\linewidth]{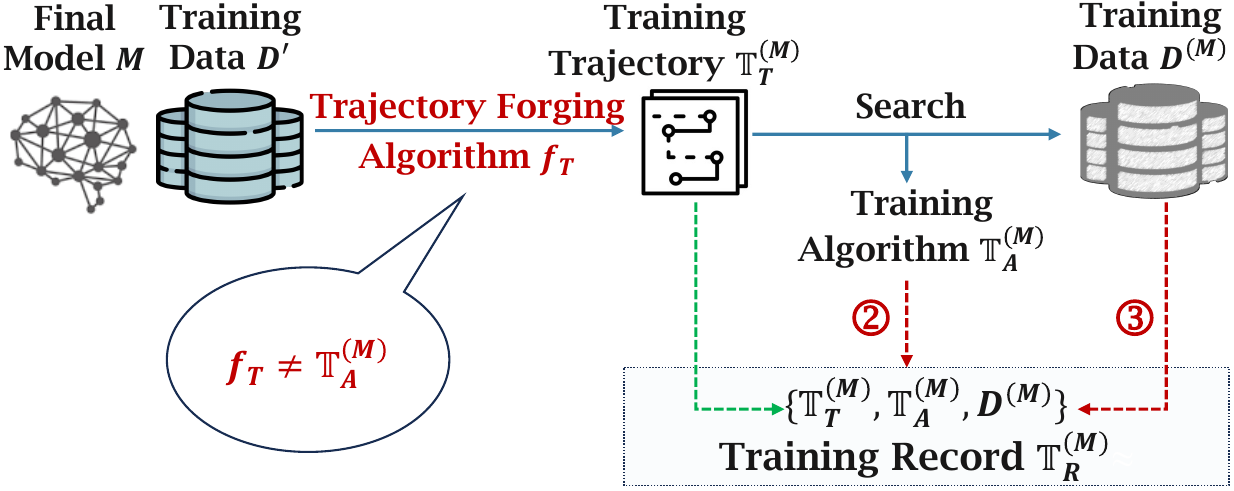}
		\caption{Reverse-Direction Attacks}
		\label{fig:model-forging}
	\end{subfigure}
        
	\caption{An illustration of four types of forging training records, where initialization algorithm $\mathbb{I}_A$ is omitted since the initial models are assumed to be honest in all cases. (a) Honest DNN training. The training trajectory is generated by operating training algorithms over training data. (b) Forward-direction attacks. Similar to honest DNN training, the training trajectory is forged after the training algorithms and data are determined. Along this direction, we depict the algorithm manipulation attacks by the red dashed line marked with {\color{red}\ding{172}}, where $f_T$ is a manipulated algorithm that utilizes final model $M$ to output the trajectory with less effort. (c) Reverse-direction attacks. After generating trajectory through $f_T$, instead of putting $f_T$ in the training record as the training algorithms, it searches training algorithms and training data so that the retraining from them can successfully lead to the trajectory. According to whether the training data are honestly sampled from the target data distribution or not, we depict the algorithm manipulation attacks and data fabrication attacks by the red dashed lines marked with {\color{red}\ding{173}} and {\color{red}\ding{174}}, respectively.}
	\label{fig:taxonomy2}
\end{figure*}

\begin{itemize}
    \item One is the (forward or reverse) direction of forging training record. As illustrated in Figure \ref{fig:taxonomy2}, the forward-direction attacks follow the same direction of outputting training records as the honest training, i.e., forging training data and algorithms at first and then generating the training trajectory by executing forged training algorithms over training data. In contrast, the reverse-direction attacks forge the training trajectory at first, and then generate the training data and algorithms in the training record. This implies that along the reverse direction, the algorithms of forging trajectory are not the training algorithms that are put in the training record.
    \item The other is the forgery object. If the data are sampled from the target data distribution as honest provers do, then the malicious prover can only manipulate training algorithms to forge the training record. We call this type of attacks \textit{algorithm manipulation attack}. Otherwise, we call it \textit{data fabrication attack}, as the data are fabricated by the malicious prover.
\end{itemize}

By the above two perspectives, we classify potentially infinite attacks into four types, as shown in Table \ref{tab:taxonomy}. Next, we model the key features of each attack type, respectively. 
\begin{table}[htbp]
\caption{Taxonomy of non-trivial attacks and their instances.}
\label{tab:taxonomy}
\centering
\resizebox{\columnwidth}{!}{%
\begin{tabular}{|c|c|c|}
\hline
\multicolumn{1}{|l|}{\textbf{}} & \textbf{Honest Data Sampling} & \textbf{Dishonest Data Sampling} \\ \hline
\textbf{\begin{tabular}[c]{@{}c@{}}Forward\\ Direction\end{tabular}} & \begin{tabular}[c]{@{}c@{}}Forward Algorithm\\ Manipulation (Attack 1)\end{tabular} & \begin{tabular}[c]{@{}c@{}}Forward Data\\ Fabrication\end{tabular} \\ \hline
\textbf{\begin{tabular}[c]{@{}c@{}}Reverse\\ Direction\end{tabular}} & \begin{tabular}[c]{@{}c@{}}Reverse Algorithm\\ Manipulation (Attack 2)\end{tabular} & \begin{tabular}[c]{@{}c@{}}Reverse Data\\ Fabrication (Attacks 3,4)\end{tabular} \\ \hline
\end{tabular}%
}

\end{table}

\subsubsection{Forward Direction Attacks}
As illustrated in Figure \ref{fig:taxonomy2}(b), the forward direction attacks consist of two sequential steps. The first step is to determine the sequence of training algorithms $\mathbb{T}_A^{(M)}$ and data $D^{(M)}$ in the training record. The second step is to generate the training trajectory $\mathbb{T}_T^{(M)}$ by running the training algorithms over training data. Hence, the effort of the malicious prover is the cost (e.g., running time) of the initialization algorithm $\mathbb{I}_A^{(M)}$ and the training algorithms $\mathbb{T}_A^{(M)}$. Recall that the goal of the malicious prover is to make its effort less than that of the honest prover. Since the initialization algorithm is assumed to be honest, the common idea of forward direction attacks is to manipulate $\mathbb{T}_A^{(M)}$ so that $Cost(\mathbb{T}_A^{(M)},\{\mathbb{T}_T^{(M)},D^{(M)}\})<Cost(\mathbb{T}_A^{(H)},\{\mathbb{T}_T^{(H)},D^{(H)}\})$. On the contrary, the malicious prover has few motivations to fabricate the training data, because there is a higher risk of being detected but no benefit in doing so. Therefore, along the forward direction, we focus on algorithm manipulation attacks.

By Assumption \ref{assu:strong}, the malicious prover cannot find manipulated algorithms with less cost unless the forged training algorithms utilize the target model for model updating. Hence, the forward direction algorithm manipulation attacks mainly consider how to integrate the target model into training algorithms to reduce the cost. We provide an example below.

\noindent\textbf{Attack 1 (forward direction)~\cite{liuProvenanceTrainingTraining2023}:} Given a target model $M$, the malicious prover honestly samples an initialized model $M_1$ and training data.
Then it generates a training trajectory $\mathbb{T}_T^{(M)}=\{M_1,\ldots,M_m\}$ by optimization from $M_1$ to $M_m$. In particular, the malicious prover generates $M_i$ from $M_{i-1}$ ($2\leq i\leq m$) by minimizing the following loss function:
\begin{equation}
    \mathcal{L}^\prime=\mathcal{L}(D_i)+\alpha\cdot d(M_i,M),
\end{equation}
where $\mathcal{L}(\cdot)$ is an honest loss function for DNN training (e.g., cross-entropy loss or mean square error loss) over dataset $D_i$, $d(M_i,M)$ is the distance between $M_i$ and $M$, and $\alpha$ is a weight coefficient of loss function.

Through this algorithm design, the models will converge to $M$ more quickly with a larger value of $\alpha$ and the honest training algorithm is in fact a special case with $\alpha=0$. Let $E$ and $t$ (resp. $E^\prime$ and $t^\prime$) denote the number of epochs and the running time of each epoch for honest (resp. forged) training algorithms. By setting a large value of $\alpha$, $E^\prime$ can be far less than $E$. Although the running time of each epoch is slightly higher (i.e., $t^\prime>t$) due to the extra term $d(M_i,M)$ in the loss function, the total cost of forged algorithms $E^\prime\cdot t^\prime$ can be lower than the total cost of honest algorithms $E\cdot t$, as long as the value of $\alpha$ is large enough.

We note that existing PoT scheme~\cite{jiaProofofLearningDefinitionsPractice2021a} treats training algorithms that utilize the target model $M$ as structure incorrect algorithms and thus excludes them from consideration. Nonetheless, since we model each algorithm as an executable program (see Section \ref{sec:modeling-record}), it is hard to detect its internal mechanism and realize its abnormality. Although the verifier can also ask the prover to submit training algorithms in the form of pseudo-code, then the verification of training algorithms requires to implement all of these algorithms by programming. Due to the reproducibility hardness of training algorithms, this way would put too much burden on the verifier and may be infeasible in practice.
Therefore, this paper assumes the algorithms are submitted in the form of executable programs and takes Attack 1 into consideration.

Next, we model the key feature of (forward direction) algorithm manipulation attacks.
Since the manipulated training algorithms must use the target model for model updating, the output of each manipulated training algorithm $\mathbb{T}^{(M)}_{A,i}$ (i.e., $\Delta_i$) has the tendency towards leading $M_{i+1}=M_i+\Delta_i$ to be close to $M$, irrespective of the training data $D_i$ and the model $M_i$ are. In contrary, the honest training algorithm $\mathbb{T}^{(H)}_{A,i}$ does not have such tendency. This indicates that the output of $\mathbb{T}^{(H)}_{A,i}$ has more diversity or uncertainty compared with $\mathbb{T}^{(M)}_{A,i}$. Such uncertainty can be measured by information entropy, with higher information entropy indicating a higher degree of uncertainty. Specifically, given a model state $M_{i}$, the entropy of $\mathbb{T}^{(H)}_{A,i}(D_i,M_i)$ is greater than or equal to that of $\mathbb{T}^{(M)}_{A,i}(D_i,M_i)$. This can be formalized as 
    \begin{equation}
        H(\mathbb{T}^{(H)}_{A,i}(D_i,M_i)\mid M_{i})\geq H(\mathbb{T}^{(M)}_{A,i}(D_i,M_i)\mid M_{i}), 
    \end{equation}
    where $H(y|x)$ denotes the entropy of random variable $y$ conditioning on the knowledge of random variable $x$.

\subsubsection{Reverse Direction Attacks}

As illustrated in Figure \ref{fig:taxonomy2}(c), the reverse direction attacks consist of the same two sequential steps as forward direction attacks but in a different order. Specifically, the first step is to forge the training trajectory $\mathbb{T}_T^{(M)}$, while the second step is to forge the sequence of training algorithms $\mathbb{T}_A^{(M)}$ and data $D^{(M)}$. We use $f_T$ and $f_A$ to denote two forging algorithms in these two steps, respectively. Notably, the trajectory is generated by $f_T$ rather than the training algorithms claimed in training record, i.e., $f_T\neq \mathbb{T}_A^{(M)}$, which is a key difference between forward and reverse direction attacks.
Next, we discuss two types of reverse attack methods that take training algorithms and training data as forgery objects, respectively.




\noindent\textit{\textbf{Reverse Algorithm Manipulation Attacks.}}
This type of attacks manipulate training algorithms $\mathbb{T}_A^{(M)}$ rather than forging training data $D^{(M)}$. The main challenge here is to guarantee that $D^{(M)}$ are honest, or more specifically, $D^{(M)}$ are sampled from the target distribution and the trajectory $\mathbb{T}_T$ can be generated by running $\mathbb{T}_A^{(M)}$ over $D^{(M)}$.
To tackle this challenge, the only meaningful method is to manipulate $\mathbb{T}_A^{(M)}$ so that its output is almost a constant with respect to input data. Otherwise, for any reverse algorithm manipulation attack, there exists a forward algorithm manipulation attack whose cost is lower. To see why, we recall that the cost of a forward attack is $Cost(\mathbb{I}_A^{(M)},\mathsf{input}_1)+Cost(\mathbb{T}_A^{(M)},\mathsf{input}_2)$, where $\mathsf{input}_1$ and $\mathsf{input}_2$ are the inputs of $\mathbb{I}_A^{(M)}$ and $\mathbb{T}_A^{(M)}$, respectively. On one hand, since we assume that the initial model is honestly generated, the algorithm $f_T$ must execute the initialization algorithm $\mathbb{I}_A^{(M)}$ and thus
\begin{equation}
    Cost(f_T,\mathsf{input}_{f_T})\geq Cost(\mathbb{I}_A^{(M)},\mathsf{input}_1),
    \label{eq:argument-1}
\end{equation}
 where $\mathsf{input}_{f_T}$ denotes the input of $f_T$.
On the other hand, by Assumption \ref{assu:strong-2}, to ensure that $\mathbb{T}_A^{(M)}$ and $D^{(M)}$ can induce $\mathbb{T}_T$, the cost of the malicious prover is at least $Cost(\mathbb{T}_A^{(M)},\mathsf{input}_2)$. Hence, we have
\begin{equation}
    Cost(f_A,\mathsf{input}_{f_A})>Cost(\mathbb{T}_A^{(M)},\mathsf{input}_2),
    \label{eq:argument-2}
\end{equation}
  where $\mathsf{input}_{f_A}$ denotes the input of $f_A$.
By combining Eq.$\eqref{eq:argument-1}$ and Eq.$\eqref{eq:argument-2}$, we have
\begin{equation} 
\begin{split}
    Cost(f_T,&\mathsf{input}_{f_T})+Cost(f_A,\mathsf{input}_{f_A})>\\
&Cost(\mathbb{I}_A^{(M)},\mathsf{input}_1)+Cost(\mathbb{T}_A^{(M)},\mathsf{input}_2).
\end{split}
\end{equation}
This indicates that it would be more valuable to conduct forward algorithm manipulation attack due to lower cost.

According to the above arguments, the only reverse attack with a lower cost than its forward counterpart is to set the output of $\mathbb{T}_A^{(M)}$ to a constant with respect to its input data. An example of such attacks is described as follows. Particularly, existing work~\cite{fangProofofLearningCurrentlyMore2023} shows that it can break existing retraining-based PoT schemes~\cite{jiaProofofLearningDefinitionsPractice2021a}.

\noindent\textbf{Attack 2 (reverse algorithm manipulation)~\cite{fangProofofLearningCurrentlyMore2023}:} Given a target model $M$, we describe the forging algorithms $f_T$ and $f_A$, respectively. Regarding $f_T$, the malicious prover honestly samples an initialized model $M_1$ and randomly generates $M_{i+1}$ with $d(M_{i+1},M_i)<B$ and $d(M_{i+1},M)<d(M_i,M)$ for $i=2,\ldots,m-1$, where $d$ denotes the distance between two model states and $B$ is the upper bound of noise $z_i$. As for $f_A$, it samples training data from the target distribution and manipulates training algorithms by setting their learning rates to be extremely small. In this way, no matter what the training data are, the output of training algorithm $\mathbb{T}_{A,i}$ (i.e., $\Delta_i$) has a small value that is close to zero. Since the output has nearly no uncertainty, its entropy is also close to zero. Meanwhile, we note that $d(M_{i+1},M_i)<B$ and thus $M_{i+1}=M_i+\Delta_i+z_i$ for some $B$-bounded noise $z_i$. Therefore, this attack can cheat retraining-based PoT schemes.

This attack has a relatively low cost. The forging algorithms $f_T$ and $f_A$ basically generate random values within some range, without expensive operations such as gradient descent. As a result, its cost is much lower than that of honest training.

\noindent\textit{\textbf{Reverse Data Fabrication Attacks.}}
Considering the difficulty of ensuring honest training data, an easier strategy is to waive the generation of training data. Instead, the forging algorithm $f_A$ simply puts the honest training algorithm (say $\mathbb{T}_A^{(H)}$) in the training record and does not generate training data.
Although such a strategy reduces the cost of $f_A$ to nearly zero, it can NOT guarantee the existence of corresponding training data, let alone that the training data match the target distribution $\mathcal{D}$. Hence, these attacks are called data fabrication attacks and are usually used when the verifier is forbidden to collect training data. In this scenario, the malicious prover primarily focuses on devising a forging algorithm $f_T$ to manipulate the trajectory. Notably, the effort of the malicious prover is mainly the cost of $f_T$, as the cost of $f_A$ is nearly zero. Below we provide two examples of attacks and discuss their costs.

\noindent\textbf{Attack 3 (reverse data fabrication)~\cite{liuProvenanceTrainingTraining2023,zhangAdversarialExamplesProofofLearning2022}:} Given a target model $M$, the malicious prover honestly samples an initialized model $M_1$ and generates a training trajectory $\mathbb{T}_T=\{M_1,\ldots,M_m\}$ as $M_i=(1-\alpha_i)M_1+\alpha_i M$ for $1\leq i\leq m$, where $\alpha_i$ gradually increases from $0$ to $1$. The cost of this attack is the running time of computing multiple linear combinations of $M_1$ and $M$. Note that computing one linear combination takes much less time than honest training in one epoch. Hence, the cost of this attack is smaller than that of honest training.

\noindent\textbf{Attack 4 (reverse data fabrication)~\cite{liuProvenanceTrainingTraining2023}:} Given a target model $M$, the malicious prover honestly samples an initialized model $M_1$ using the same random seed as the honest prover and generates trajectory $\mathbb{T}_T=\{M_1,\ldots,M_m\}$ by optimization from $M_m$ to $M_1$. Concretely, the malicious prover sets $M_m=M$ and generates $M_{i-1}$ from $M_i$ ($3\leq i\leq m$) by minimizing the loss function:
\begin{equation}
    \mathcal{L}^\prime=\mathcal{L}(D^{false}_i)+\alpha\cdot d(M_i,M_1),
\end{equation}
where $D_i^{false}$ is a false dataset in which some labels are set to be wrong, $\mathcal{L}(\cdot)$ is a basic loss function for DNN training (e.g., cross-entropy loss or mean square error loss) over dataset $D_i^{false}$, $d(M_i,M_1)$ is the distance between $M_i$ and $M_1$, and $\alpha$ is a weight coefficient of loss function.

The cost of this attack is the product of the number of epochs $E^\prime$ and the running time of each epoch $t^\prime$, i.e., $E^\prime\cdot t^\prime$.
By the algorithm design, the models will converge to $M_1$ more quickly with a larger value of $\alpha$. Hence, by setting a large value of $\alpha$, the value of $E^\prime$ can be small enough to make the cost of this attack less than the honest prover's cost.

\subsubsection{A Successful Attack against Statistics-Based PoT}
\label{sec:successfulattack}
Next, we show a successful attack against the existing statistics-based PoT scheme~\cite{liuProvenanceTrainingTraining2023} in a more realistic setting.
On one hand, the existing scheme assumes that the malicious prover only has a small dataset and is not aware of the initial model of honest training trajectory, which is different from our threat model (see Section \ref{sec:modeling-threat}).
On the other hand, the existing scheme assumes that the prover will save model states in every epoch, which may result in excessive storage cost. Hence, in practice, the model owner usually save model states in some checkpoints, and there are usually several epochs between these two checkpoints. We set the number of epochs between two checkpoints to be $5$ in the following attack.

In the new setting, we find that the existing scheme~\cite{liuProvenanceTrainingTraining2023} may fail to distinguish honest training records from forged ones. Specifically, existing scheme examines six properties of training trajectory (labeled from $1$ to $6$) to distinguish between honest and forged training records. We perform honest training and Attack $4$ to output an honest trajectory and a forged trajectory upon CIFAR10 dataset and ResNet18 model architecture. Then we run the existing scheme to examine properties $1$--$6$ on these two trajectories.
The results demonstrate that the honest trajectory and the forged trajectory show the same value on every properties except property $2$. The property $2$ measures the maximum distance between two model states in successive checkpoints. They claim that the honest training record should have a smaller value on this metric.
However, in our experimental results, the metric of the honest training record ($0.021$) is greater than that of malicious training record ($0.019$). Therefore, the existing scheme fail to distinguish between the honest prover and malicious provers in this case.

\section{Criterion for Attack Detection}
\label{sec:criterion}

Based on the above modeling, we identify two types of attack methods (namely algorithm manipulation and data fabrication attacks) beyond those attacks that can be defended by existing PoT schemes. Still, we are not aware of how to detect these attacks. In this section, we provide a universal criterion for detecting these attacks. Next, we first provide a criterion for detecting algorithm manipulation attacks and then demonstrate that this criterion is also effective for detecting data fabrication attacks.

\noindent\textbf{\textit{Algorithm Manipulation Attack Detection.}}
To detect algorithm manipulation attacks, we need to check whether algorithms in the training record are manipulated. However, since we assume training algorithms are submitted in the form of executable programs, it is hard to directly examine whether the algorithms are manipulated. Instead, we can only check the inputs and outputs of training algorithms. Below we analyze the distinctions between honest and manipulated training algorithms, in terms of the relationships between their inputs (i.e., training data) and outputs (i.e., training trajectory).

Intuitively speaking, the training trajectory from manipulated algorithms should be \textit{less} dependent on the training data, as the manipulated algorithms utilize additional values beyond training data (e.g., the final model) to generate the trajectory.
To formally prove this intuition, we employ information theory~\cite{shannon} and adopt mutual information to measure the dependence between training trajectory and training data. The mutual information between two random variables $x$ and $y$ is denoted by $I(x;y)$, with its larger value representing stronger dependence. By Theorem \ref{theo:dis}, we claim that compared with manipulated training algorithms, the trajectory from honest training algorithms should be more dependent on the input training data. For space limitation, the proof of Theorem 1 could be referred to Appendix \ref{appendix-proof}.

\begin{thmprime}{theo:dis}

 Suppose that the malicious prover conducts algorithm manipulation attacks with a lower cost than honest training. Let $(\mathbb{I}_A,\mathbb{T}_A,\mathbb{T}_T,D)$ and $(\mathbb{I}_A,\mathbb{T}_A^{(M)},\mathbb{T}_T^{(M)},D^{(M)})$ denote the training records of the honest prover and the malicious prover, respectively.  When Assumptions \ref{assu:strong}-\ref{assu:strong-2} hold, we have    
    \begin{equation}
        I(D;\mathbb{T}_{T,[i:i+k]} )> I(D^{(M)};\mathbb{T}_{T,[i:i+k]}^{(M)}),
    \end{equation}
    where $\mathbb{T}_{T,[i:i+k]}$ and $\mathbb{T}_{T,[i:i+k]}^{(M)}$ are two length-$k$ fragments from $\mathbb{T}_T$ and $\mathbb{T}_{T}^{(M)}$, respectively.
    
\end{thmprime}

Inspired by Theorem \ref{theo:dis}, a direct solution for manipulated algorithm detection is to measure the mutual information between training trajectory and data: the smaller the value of mutual information, the more likely it is that the training algorithms are manipulated. However, this solution may be impractical because the dimension of training trajectory is too large. Taking ResNet-18 as an example, each model state is a 270000-dimensional vector, and the trajectory typically consists of hundreds or even thousands of model states. Directly measuring the mutual information between this high-dimensional vector and others is usually impractical.

\begin{figure}[tbp]
	\centering
	\begin{subfigure}{0.49\linewidth}
		\centering
		\includegraphics[width=\linewidth]{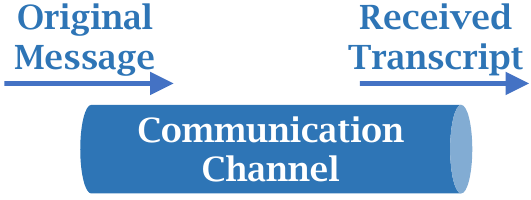}
		\caption{Communication Systems}
		\label{channel1}
	\end{subfigure}
	\centering
	\begin{subfigure}{0.49\linewidth}
		\centering
		\includegraphics[width=\linewidth]{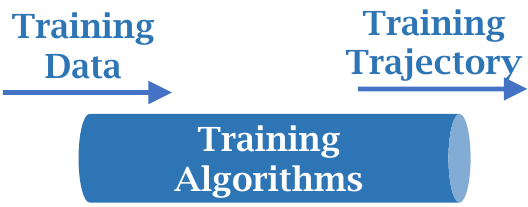}
		\caption{DNN Training}
		\label{channel2}
	\end{subfigure}
        \caption{Analogy between communication and training.}
	\label{channel}
\end{figure}

To tackle the above challenge, we find another way to understand the implication of mutual information in this scenario by learning from the application of information theory in communication systems. In a communication system, a message is transmitted from a sender to a receiver through a channel (see Figure \ref{channel1}). By Shannon's channel coding theorem~\cite{shannon}, for a channel, the mutual information between its input (original message) and output (received transcript) measures the optimal capability to recover the original message from the received transcript.
The greater the value of such mutual information, the more likely it is that the receiver can recover the original message from its received transcript.
Here, as illustrated in Figure \ref{channel2}, the training algorithms are like a channel, with its input being training data and its output being training trajectory. Correspondingly, by Theorem \ref{theo:dis}, compared with an honest training record, it should be harder to recover training data from the training trajectory in a malicious training record due to its smaller mutual information.

Based on the above discussion, we propose a criterion to detect algorithm manipulation attacks using training algorithms $\mathbb{T}_A$ and trajectory $\mathbb{T}_T$. The key insight is to see whether we can recover \textit{high-fidelity} training data from the training algorithms and trajectory. More precisely, we try to find a synthetic data $S$ so that operating training algorithms $\mathbb{T}_A$ over $S$ induces a similar trajectory as $\mathbb{T}_T$. The fidelity of synthetic data (i.e., whether synthetic data are close to the target data distribution) measures the effect of recovering the training data from the training trajectory.
Hence, we can detect algorithm manipulation attacks based on the fidelity of synthetic data.
The higher the fidelity of synthetic data, the more likely it is that the training record is from the honest prover.

Now we formally state the criterion for detecting algorithm manipulation attacks. That is, if the verifier finds synthetic data $S$ that can support approximate retraining (i.e., executing training algorithms over $S$ leads to a similar training trajectory), then the optimal fidelity of synthetic data $S$ from malicious provers is lower than that from honest provers. In the next subsection, we show that this criterion is universal for detecting other attacks.

\noindent\textbf{\textit{Data Fabrication Attack Detection.}}
Next, we demonstrate that the above criterion can also be used to detect data fabrication attacks.
Recall that the key feature of data fabrication attacks is its dishonest training data, i.e., either there is no training data that can generate the training trajectory via the training algorithms, or the training data will deviate far from the target data distribution. In this case, if the verifier tries to find training data $S$ that can support approximate retraining, the best it can do is that $S$ will deviate far from the target distribution. In contrast, for the training record from the honest prover, the optimal $S$ is the real training data $D$, which are honestly sampled from the target distribution. Hence, if the verifier finds synthetic data $S$ that can support approximate retraining, the synthetic data with higher fidelity are unlikely to perform data fabrication attacks. 

Based on the above reasoning, the criterion proposed to detect algorithm manipulation attacks can also be used to detect data fabrication attacks. Therefore, by utilizing this criterion, we can detect all attacks at once, rather than detect each attack one by one.

\section{Our Generic PoT Construction}
\label{sec:scheme}

In this section, we show how the universal criterion can guide us to design secure PoT schemes. Before presenting our PoT construction, we first describe the basic settings on verifier capability that follows the existing PoT scheme~\cite{liuProvenanceTrainingTraining2023}. On one hand, we assume that the verifier is honest, i.e., it will faithfully execute the verifying algorithm. Still, considering existing laws and regulations about data privacy, our PoT scheme will not send training data to the verifier. On the other hand, we assume that the verifier has a test dataset $D_{test}$ with a similar distribution as the training data. In practice, the test dataset can be obtained from public dataset or be selected from those data on which the target model has high confidence to output correct inference results.

Next, we present our PoT construction under the above setting. Generally speaking, a PoT scheme is a pair $(\mathcal{P},\mathcal{V})$, where $\mathcal{P}$ is a \textit{proving algorithm} that takes a training record as input and outputs a proof, and $\mathcal{V}$ is a \textit{verifying algorithm} that takes multiple proofs as input and selects a proof as the one from honest provers. In our PoT construction, the proving algorithm is to simply put training record except training data (i.e., $\{\mathbb{I}_A,\mathbb{T}_A,\mathbb{T}_T\}$) into the proof. For the verifying algorithm, based on the modeling of attack methods, the verifying algorithm needs to check the following items.
\begin{enumerate}
    \item Is the proof structurally correct?
    \item Is the model initialization honest?
    \item Does the prover conduct algorithm manipulation or data fabrication attacks?
\end{enumerate}

Among these items, the former two items can be checked by following existing PoT schemes~\cite{jiaProofofLearningDefinitionsPractice2021a,gargExperimentingZeroKnowledgeProofs2023}, while the third item can be examined based on the aforementioned universal criterion. Correspondingly, we decompose the verifying algorithm into three stages. Stage $1$ is used to check the former two items, while Stages $2$ and $3$ are performed to detect attacks according to the criterion. More specifically, Stage $2$ synthesizes the data that induce similar trajectory with the same training algorithms and Stage $3$ examines the fidelity of synthetic data. Below we present the concrete design of these three stages, respectively.

\textbf{Stage 1: Structure and Initialization Check.} The verifier first checks whether the training record matches the requirements of correct structure. If so, the verifying algorithm checks whether the initialized model is honest. There are two possibilities. 
If the initialization algorithm $\mathbb{I}_A$ is to sample a random model from some distribution, the verifying algorithm adopts Kolmogorov–Smirnov test~\cite{massey1951kolmogorov} to check whether $M_1$ is truly sampled from distribution specified in $\mathbb{I}_A$. If the initialization algorithm $\mathbb{I}_A$ is to use an existing well-trained model, the verifying algorithm checks whether this well-trained model has corresponding honest training record.

\textbf{Stage 2: Data Synthesis.} By taking training algorithms and training trajectory as input, the verifying algorithm generates a synthetic data $S$ that can support approximate retraining (i.e., training on $S$ leads to a similar trajectory as training on original training data $D$). To do so, we borrow the trajectory matching algorithms from researches on data distillation.

Given a training data $D$, data distillation aims to create a small, synthetic data $S$ from $D$ such that a model trained on $S$ will have similar accuracy as a model trained on $D$. Trajectory matching is one of data distillation algorithms. Its key idea is to train a teacher model on $D$ and a student model on $S$, and to encourage the consistency of trajectories of these two models.
To adopt trajectory matching algorithms in our application, the trajectory of teacher model is the training trajectory from the prover, while the verifier needs to train the student model by itself over synthetic data.
By minimizing the distance between trajectories of teacher model and student model, trajectory matching algorithms can help us to optimize $S$ to support approximate retraining. More details, including the pseudo-random code of trajectory matching algorithms, can be found in Appendix \ref{appendix-code}.


\textbf{Stage 3: Data Evaluation.} To evaluate whether synthetic data $S$ are sampled from the target data distribution, the verifying algorithm measures the dependence between $S$ and test data $D_{test}$. Among various existing methods, we choose to train neural networks to measure this dependence. More specifically, the verifying algorithm trains $t$ new models $M^{(new)}_{1},\ldots,M^{(new)}_{t}$ using synthetic data $S$ and tests the accuracy of these models over test dataset $D_{test}$. To obtain a more accurate estimate of data fidelity, we train these new models starting from different initialized models and compute the average accuracy across these $t$ new models. 
Among multiple proofs from different provers, the verifying algorithm outputs the proof whose average accuracy is highest as the proof from the honest prover.

Based on the above construction, we can instantiate a concrete PoT scheme by specifying the details of these stages, such as the adopted trajectory matching algorithm and the setting of related parameters. Hence, the development of trajectory matching algorithms can help to advance the PoT scheme design, indicating an intersection of independent interests within both research lines.

\section{Experiment Evaluation}
\label{sec:experiment}
In this section, we implement a prototype of our PoT scheme design and evaluate its security through experiments. The experiment setup is described in Section \ref{sec:exp-setep} and the results are shown in Section \ref{sec:exp-results}.
\begin{figure*}[htbp]
	\centering
	\begin{subfigure}{0.33\linewidth}
		\centering
		\includegraphics[width=\linewidth]{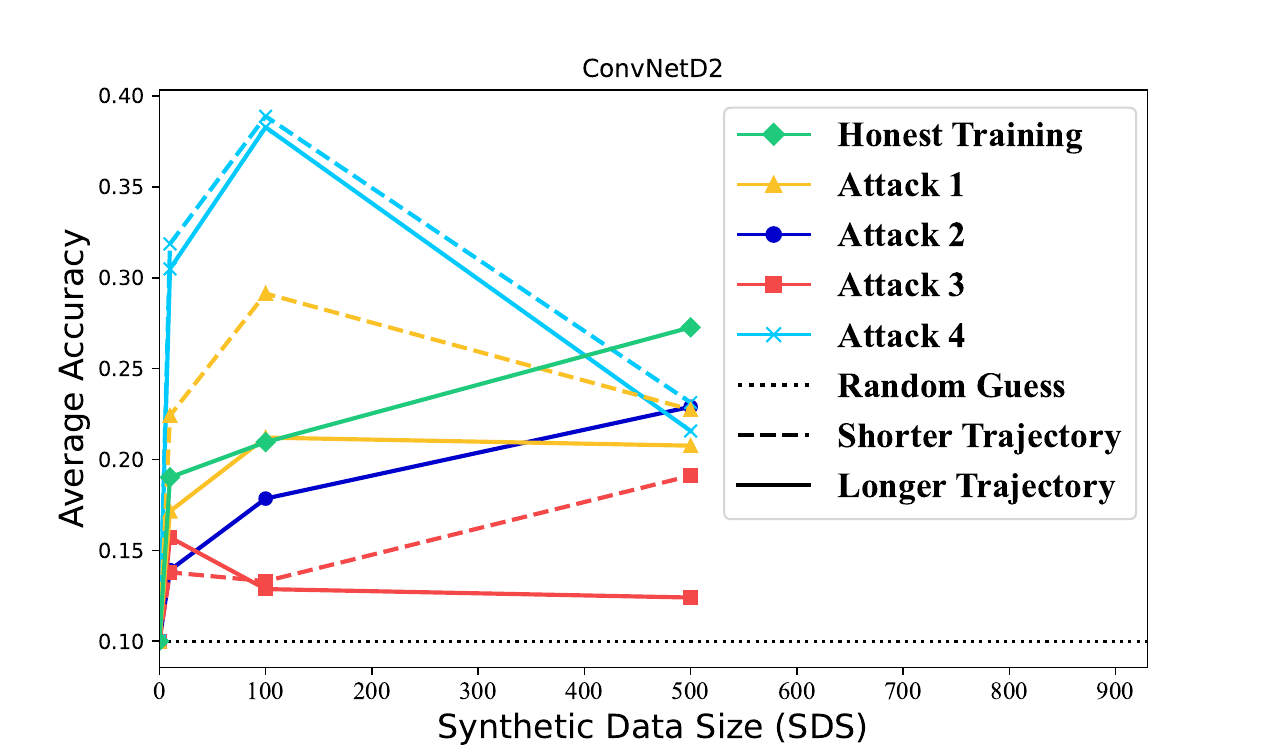}
		\caption{CIFAR10, CNN}
		\label{CIFAR10 CNN}
	\end{subfigure}
	\centering
	\begin{subfigure}{0.33\linewidth}
		\centering
		\includegraphics[width=\linewidth]{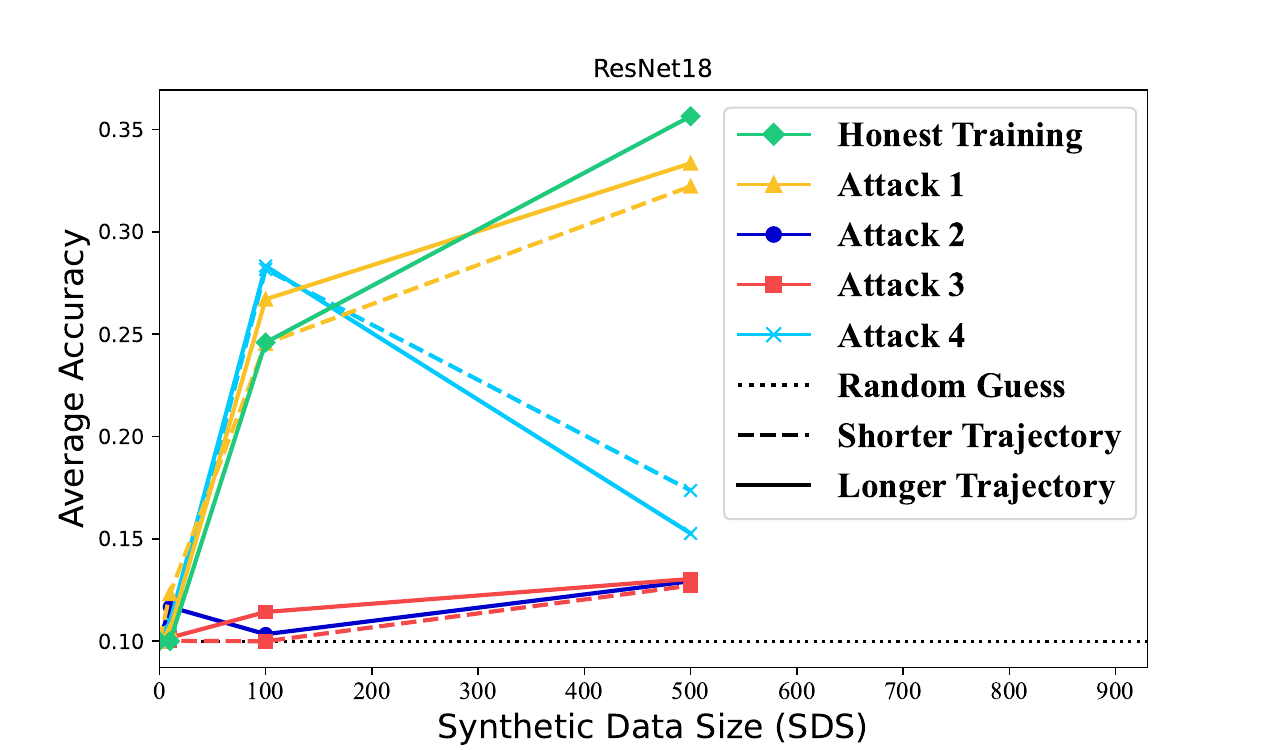}
		\caption{CIFAR10, ResNet18}
		\label{CIFAR10 ResNet18}
	\end{subfigure}
	\centering
	\begin{subfigure}{0.33\linewidth}
		\centering
		\includegraphics[width=\linewidth]{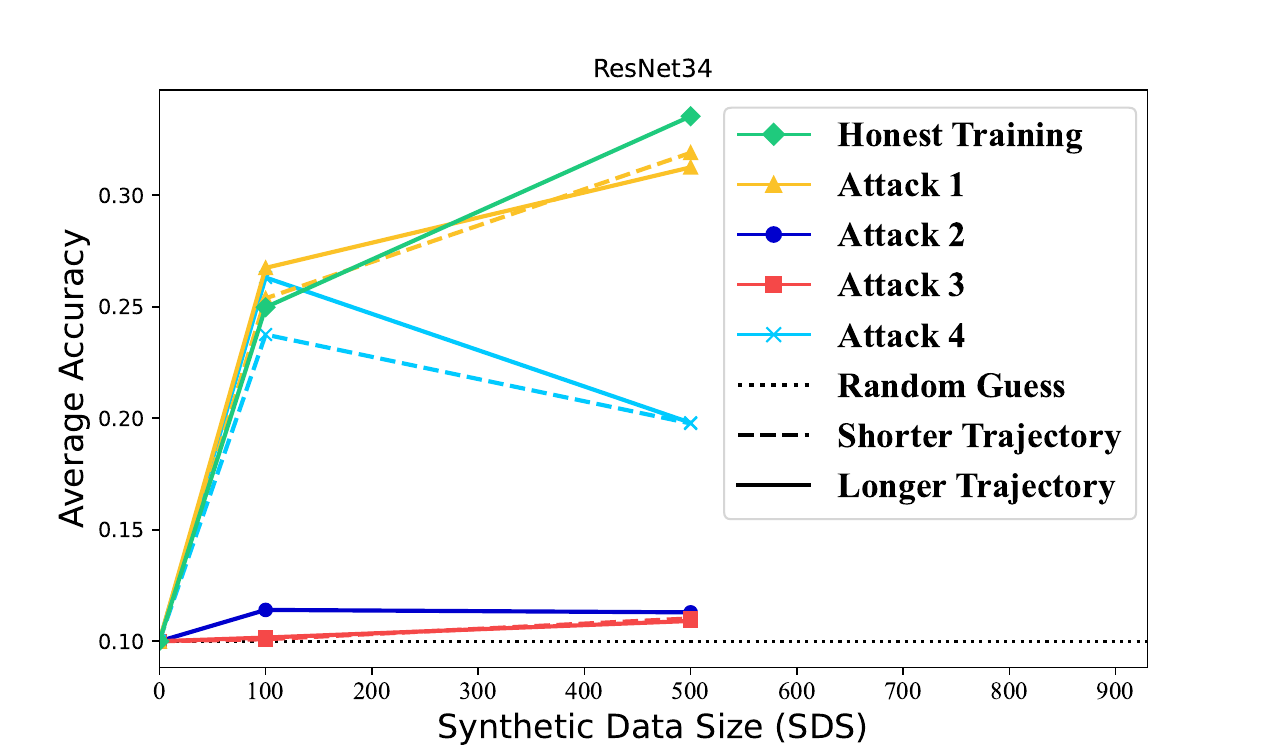}
		\caption{CIFAR10, ResNet34}
		\label{CIFAR10 ResNet34}
	\end{subfigure}
        
        \centering
	\begin{subfigure}{0.33\linewidth}
		\centering
		\includegraphics[width=\linewidth]{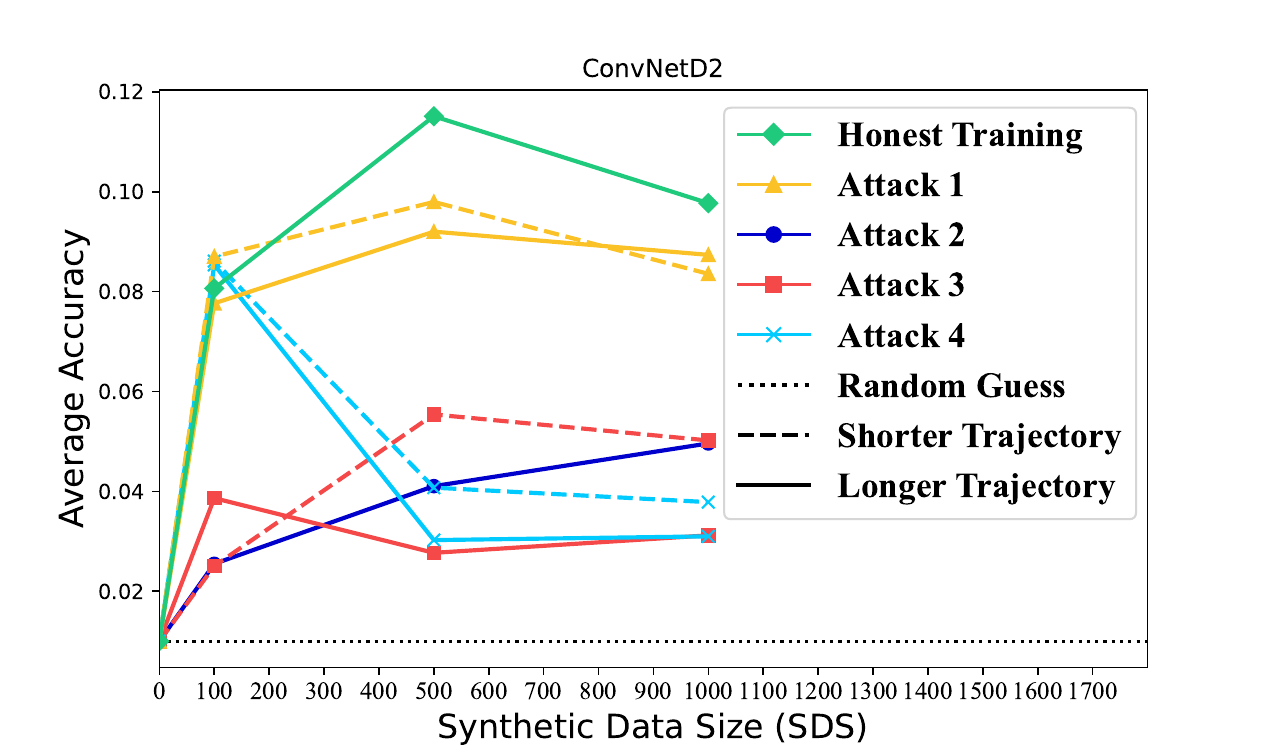}
		\caption{CIFAR100, CNN}
		\label{CIFAR100 CNN}
	\end{subfigure}
	\centering
	\begin{subfigure}{0.33\linewidth}
		\centering
		\includegraphics[width=\linewidth]{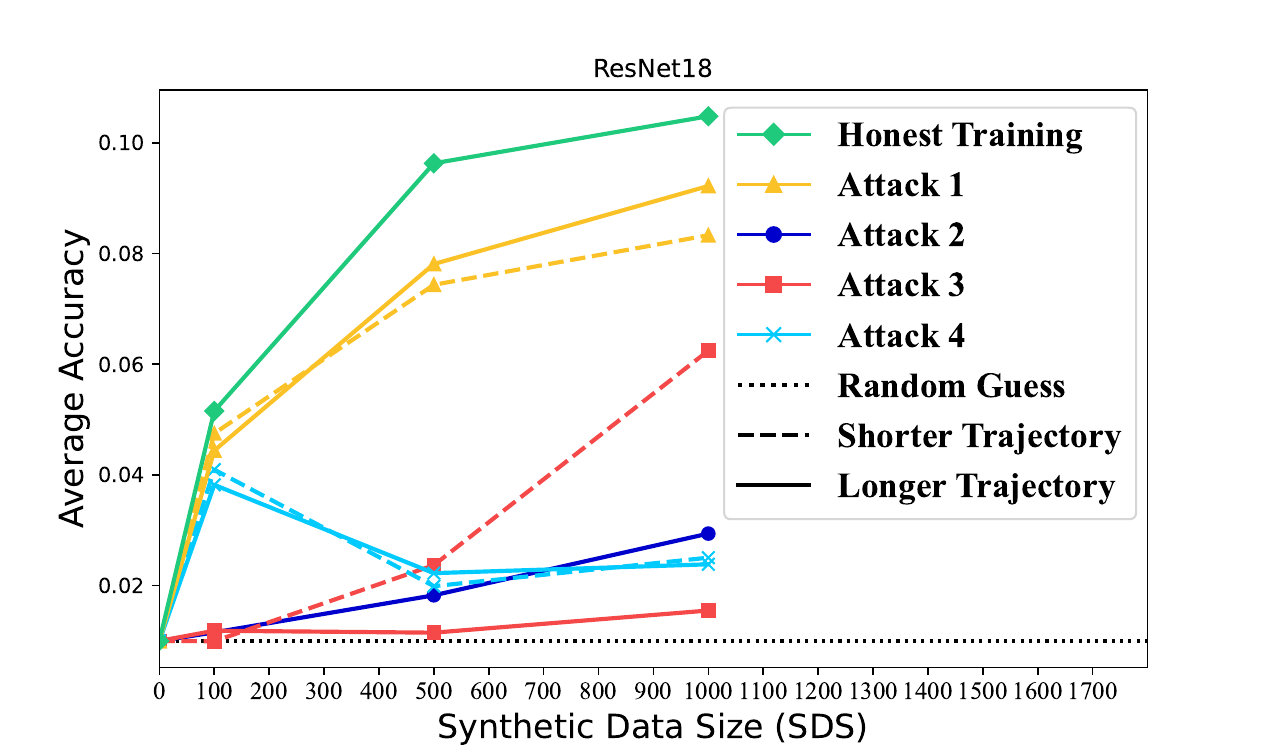}
		\caption{CIFAR100, ResNet18}
		\label{CIFAR100 ResNet18}
	\end{subfigure}
	\centering
	\begin{subfigure}{0.33\linewidth}
		\centering
		\includegraphics[width=\linewidth]{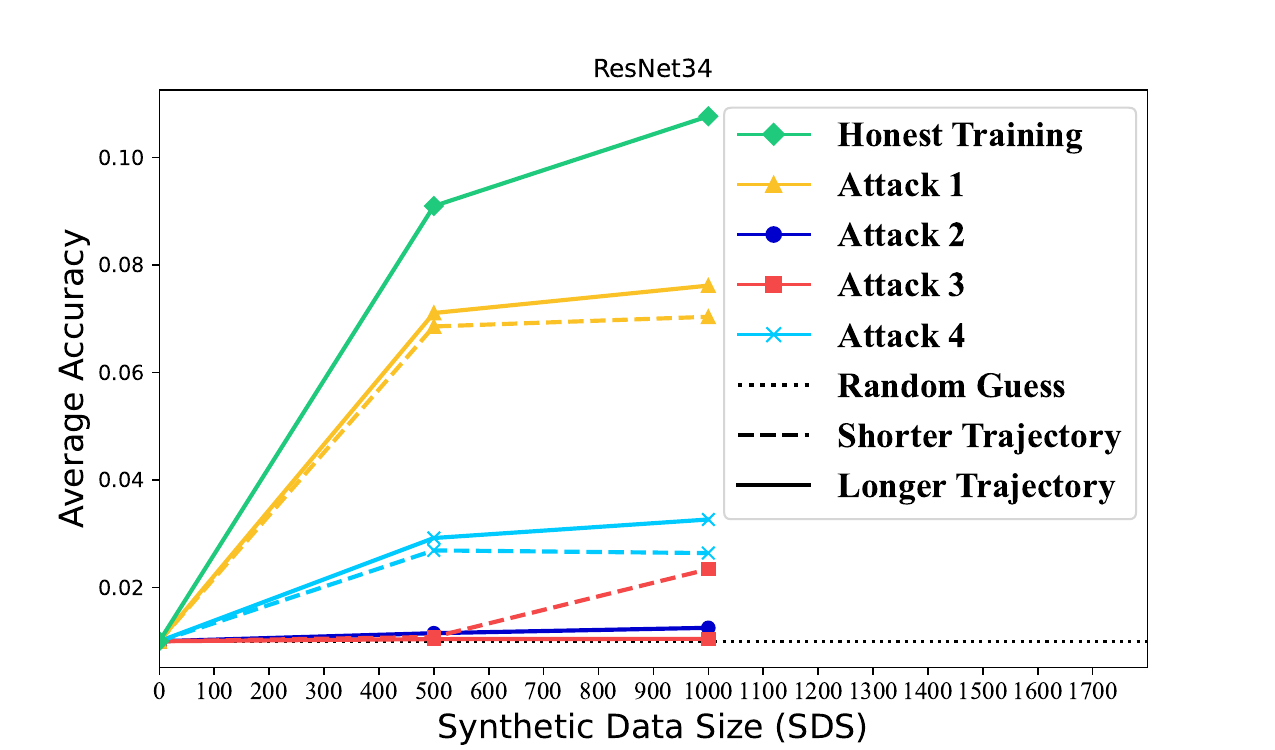}
		\caption{CIFAR100, ResNet34}
		\label{CIFAR100 ResNet34}
	\end{subfigure}
	\caption{Experimental results of average accuracy upon various model architectures, datasets, and synthetic data size (SDS).}
	\label{CNN}
\end{figure*}
\subsection{Experiment Setup}
\label{sec:exp-setep}
\noindent\textbf{Datasets and Models.} To observe the impacts of datasets and model architectures on the security of our PoT scheme, we evaluate our PoT scheme across two classical CV datasets (i.e., CIFAR10 with ten classes of images and CIFAR100 with one hundred classes of images) and three model architectures (i.e., a three-layer CNN model, a ResNet18 model, and a ResNet34 model). The training datasets and test datasets are used by provers and the verifier, respectively.

\noindent\textbf{Implementation of Attacks.} After training these models following the honest way, we simulate Attacks $1$--$4$ as described in Section \ref{sec:modeling-attack}. For Attacks $1$, $3$, and $4$, they have tunable parameters denoted by $\alpha$ or $\alpha_i$ in their descriptions, which will affect the length of output trajectory.
Specifically, for Attacks $1$ and $4$, the larger the value of $\alpha$, the more quickly the model converges and thus the shorter the trajectory. In contrast, for Attack $3$, the larger value of $\alpha_i$ implies more intermediate model states in the trajectory and hence a longer trajectory.
To observe the impact of trajectory length, we forge two training records for each of Attacks $1$, $3$, and $4$ with different values of $\alpha$ or $\alpha_i$. 
One record has a shorter trajectory and the other has a longer trajectory. In total, we forge seven training records by executing Attacks $1$--$4$.

\noindent\textbf{Implementation of PoT Scheme.} We implement our PoT scheme with Python. The stage of data distillation is implemented based on existing trajectory matching algorithm~\cite{cuiScalingDatasetDistillation2023}. To guarantee the convergence of synthetic data, this stage executes $300$ iterations so that the synthetic data are barely updated. In addition, all related parameters follow the default setting. Particularly, a critical parameter is the synthetic data size (SDS), i.e., the number of images in the synthetic data. For CIFAR10 (resp. CIFAR100) dataset, the value of SDS is set to be $10$, $100$, or $500$ (resp. $100$, $500$, or $1000$).
For the data evaluation stage, we train three DNNs (i.e., $t=3$) using the synthetic data and observe their average accuracy.


\subsection{Experiment Results}
\label{sec:exp-results}

\subsubsection{Overall Results}
We depict the evaluation results upon various model architectures and datasets in Figure \ref{CNN}. In each sub-figure, we show the average accuracy for each training record with different settings of SDS, along with a random guess line for reference (i.e. $0.1$ for CIFAR10 and $0.01$ for CIFAR100). We also note that for ResNet34 architecture, the synthetic data with SDS=10 (resp. SDS=100) is too small to converge under CIFAR10 dataset (resp. CIFAR100 dataset), with the loss being ``NAN''. Hence, we do not draw the corresponding results in Figures \ref{CIFAR10 ResNet34} and \ref{CIFAR100 ResNet34}.

From the results in Figure \ref{CNN}, we observe that the average accuracy of honest training record is highest under the maximum value of SDS, no matter what the model architecture and dataset are. This demonstrates that our PoT scheme can correctly select the honest training record among various forged training records, which corroborates the security of our PoT scheme against Attacks $1$--$4$ under different settings.

Although experimental results support the security of our PoT scheme, the defense effects are not exactly the same for different settings and attack methods. On one hand, the different settings of SDS, model architectures, and datasets lead to distinct defense effects or even different results on the decision of honest prover. On the other hand, under different attack methods, the degree of distinctions between honest training records and forged training records varies. Below we discuss the impact of these settings and attack methods on the defense effects, respectively. To measure the defense effect, we adopt the discrimination degree between honest and malicious training records as the metric in the following discussions. Specifically, given the average accuracy of honest training record (denoted by $acc_h$) and the average accuracy of malicious training record (denoted by $acc_m$), the discrimination degree $\eta$ is defined as $\eta = acc_h/acc_m$.
The larger the value of discrimination degree $\eta$, the better the defense effect. Particularly, the PoT scheme succeeds to defend against some attack if and only if $\eta>1$.

\subsubsection{Impact of Settings on Defense Effect}
To observe the impact of basic settings on the defense effect, we compare the values of discrimination degree under different settings of model architecture, dataset, and SDS. To measure the defense effects against the most powerful attack, we consider the worst-case discrimination degree, i.e., the smallest discrimination degree among all attack methods. Table \ref{tab:SDS} shows the worst-case discrimination degree in different settings of SDS, model architecture, and dataset.
\begin{table}[]
\centering
\caption{Worst-case discrimination degree in different settings of SDS, model architecture, and dataset.}
\label{tab:SDS}
\resizebox{\columnwidth}{!}{%
\begin{threeparttable}
\begin{tabular}{|l|c|c|c|c|}
\hline
\textbf{\diagbox{Model and Dataset}{SDS}} & \multicolumn{1}{c|}{10} & \multicolumn{1}{c|}{100} & \multicolumn{1}{c|}{500} & \multicolumn{1}{c|}{1000} \\ \hline
$\qquad\quad\quad$CIFAR10, CNN       & $0.60$ & $0.54$ & $1.19^{\text{+}}$ & $-$    \\ \hline
$\qquad\quad\quad$CIFAR10, ResNet18  & $0.81^{*}$ & $0.87$ & $1.07^{\text{+}}$ & $-$    \\ \hline
$\qquad\quad\quad$CIFAR10, ResNet34  & $-$    & $0.94$ & $1.09^{\text{+}}$ & $-$    \\ \hline
$\qquad\quad\quad$CIFAR100, CNN      & $-$    & $0.94$ & $1.13$ & $1.17^{\text{+}}$ \\ \hline
$\qquad\quad\quad$CIFAR100, ResNet18 & $-$    & $1.07^{*}$ & $1.16$ & $1.30^{\text{+}}$ \\ \hline
$\qquad\quad\quad$CIFAR100, ResNet34 & $-$    & $-$    & $1.26^{*}$ & $1.41^{\text{+},*}$ \\ \hline
\end{tabular}%

\begin{tablenotes}
        \footnotesize
        \item + The maximum value of discrimination degree in each row.
        \item * The maximum value of discrimination degree in each column.
      \end{tablenotes}
  \end{threeparttable}
  }
\end{table}

\noindent\textbf{The setting of SDS value.} By observing each row in Table \ref{tab:SDS}, we can tell that in almost every rows, the worst-case discrimination degree grows with the increase of SDS value, indicating that the defense effect becomes better. Recent research on trajectory matching algorithm~\cite{recentTrajectoryMatching} helps to explain this phenomenon. When the value of SDS grows, the optimization of SDS requires to extract more \textit{detailed} information about training data from the trajectory. In other words, performing the trajectory matching algorithms with the larger value of SDS helps to detect whether the trajectory contains enough detailed information. Hence, we should adopt a larger value of SDS in the PoT scheme design to improve the defense effect. However, the employment of synthetic data with larger size will simultaneously result in low efficiency. We leave the study of trade-off between defense effect and efficiency as an interesting topic in the future work.
\begin{table*}[]
\centering

\caption{Discrimination degree under different attack methods with the SDS being the maximum value.}
\resizebox{15cm}{!}{%
\begin{threeparttable}
\begin{tabular}{|l|c|c|c|c|c|c|c|}
\hline
\textbf{\diagbox{Model and Dataset}{Attack Methods}} & 1(Longer) & 1(Shorter)     &    2         & 3(Longer)                & 3(Shorter) & 4(Longer) & 4(Shorter) \\ \hline
$\qquad\quad\quad$CIFAR10, CNN                                          & $1.31$     & $1.20$     & $1.19^{*}$ & $2.2$                 & $1.43$  & $1.26$     &  $1.19^{*}$    \\ \hline
$\qquad\quad\quad$CIFAR10, ResNet18                                     & $1.07^{*}$ & $1.11$     & $2.75$ & $2.75$                 & $2.81$ &  $2.33$    & $2.05$ \\ \hline
$\qquad\quad\quad$CIFAR10, ResNet34                                     & $1.10$        & $1.09^{*}$     & $2.99$ & $3.05$                 & $3.05$  & $1.68$ &  $1.68$ \\ \hline
$\qquad\quad\quad$CIFAR100, CNN                                         & $1.17^{*}$        & $1.18$     & $1.96$            & $3.27$   &  $1.96$  &  $3.27$ & $2.58$ \\ \hline
$\qquad\quad\quad$CIFAR100, ResNet18                                    & $1.16^{*}$        & $1.26$ & $3.56$        & $6.55$   & $1.69$ & $4.37$ & $4.19$ \\ \hline
$\qquad\quad\quad$CIFAR100, ResNet34                                    & $1.41^{*}$        & $1.54$        & $9.00$        & $10.8$ & $4.70$ & $3.27$ & $4.15$ \\ \hline
\end{tabular}%
\begin{tablenotes}
        \footnotesize
        \item * The minimum value of discrimination degree in each row.
      \end{tablenotes}
  \end{threeparttable}
}

\label{tab:attack}

\end{table*}

\noindent\textbf{The setting of model architecture and dataset.} For the three model architectures, the sequence in the order of complexity from lowest to highest is CNN, ResNet18, and ResNet34. Meanwhile, the CIFAR100 dataset is more complex than CIFAR10 dataset due to more classes of images. By observing each column in Table \ref{tab:SDS}, there exists a general trend in which the defense effect becomes better upon more complex model architecture and dataset. Particularly, in each column, the maximum value of discrimination degree occurs in the setting of the most complex model architecture and dataset. This demonstrates that our PoT scheme is more applicable to large DNN models and complex datasets. Note that in practice, AI enterprises usually wish to protect the intellectual property for large DNN models trained on complex datasets, rather than small models trained on simple datasets. Hence, our PoT scheme has promising prospect for real-world applications.


\subsubsection{Impact of Attack Methods on Defense Effect}
To observe the defense effect against various attacks, we compare the values of discrimination degree under different attack methods with the SDS being the maximum value, which is shown in Table \ref{tab:attack}.

\noindent\textbf{Comparison among different attacks.} By observing each row in Table \ref{tab:attack}, we can tell that in almost every rows, the minimum value of discrimination degree occurs in Attack 1. This indicates that Attack $1$ is the strongest attack method against our PoT scheme. The reason is that Attack 1 is the most similar to honest training. The only difference between Attack 1 and honest training is the adoption of the final model in the loss function. Meanwhile, in most cases, Attacks 2 and 3 result in the maximum value of discrimination degree, i.e., they are the weakest attack method against our PoT scheme. The reason is that these two attack methods do not utilize the training data at all when forging the training trajectory. The above conclusions further confirm that our PoT scheme is effective in detecting the information of honest training data that can be extracted from the training trajectory.

\noindent\textbf{Impact of trajectory length.} Recall that for Attacks 1, 3, and 4, we forge two training trajectories with different lengths. However, in almost all cases, the discrimination degree under an attack with a shorter trajectory is similar or even identical to the discrimination degree under the same attack with a longer trajectory. This implies that modifying the length of the trajectory has little impact on the defense effect.

\section{Related Work}
\label{sec:related}
In this section, we review existing PoT schemes, which can be divided into the following two categories.


In \textit{retraining-based} PoT schemes, if the training trajectory can be generated by retraining with the same algorithm and data, the training record passes test and the ownership claim is approved. The most classical PoT scheme~\cite{jiaProofofLearningDefinitionsPractice2021a} belongs to this type. However, in this scheme, the model owner has to send the training data to the verifier. Considering that training data often contain sensitive information, this scheme can not be applied when data privacy is a key concern~\cite{liuProvenanceTrainingTraining2023}. To solve this problem, some recent works propose to adopt cryptographic primitives (e.g., zero-knowledge proofs) to conduct the retraining test without leaking the training data~\cite{sunZkDLEfficientZeroKnowledge2023,gargExperimentingZeroKnowledgeProofs2023}. Nonetheless, the adoption of cryptographic primitives leads to high communication and computation complexities with respect to the number of model parameters and the size of training data. As a result, these schemes can be used only in simple models (e.g., linear regression model) or small datasets.

In \textit{statistics-based} PoT schemes, the verifier only collects training trajectory (without training data) and checks its coherence by analyzing several statistics metrics~\cite{liuProvenanceTrainingTraining2023}. If these metrics all fall within the pre-specified range, the ownership claim is approved. Since the statistics-based category does not need to disclose training data to the verifier, it is thought to have a natural superiority on preserving data privacy.


However, regardless of the category of schemes, existing PoT schemes fail to defend against more tricky attacks.
For example, existing works develop adaptive attacks against the retraining-based PoT schemes~\cite{zhangAdversarialExamplesProofofLearning2022,fangProofofLearningCurrentlyMore2023}.
These ever-emerging attacks indicate an insufficient considerations of possible attacks. Particularly, even though existing works consider several attacks and discuss whether their PoT schemes can defend against these attacks, they fail to exclude all reasonable attacks from success. 
In this work, we provide a comprehensive modeling of attack methods and analyze their common features. Such modeling and analysis enable a goal-oriented PoT design by revealing the range of attacks that PoT should defend against.



\section{Conclusion and Discussion}
\label{sec:conclusion}
In this paper, we adopt the formal methods in the area of PoT security to devise PoT schemes with stronger security. Unlike existing works relying on intuitions or observations, we conduct theoretical modeling and analysis to identify distinctions between honest and forged training records. Following this way, we discover a universal criterion for attack detection and further propose a generic PoT construction. The empirical experiments validate the security of our PoT scheme against various attacks.
We expect this work can establish a good foundation towards understanding and enhancing the security of PoT schemes, but future work is still required.
Specifically, this work may leave the following problems that require further investigation:
\begin{itemize}
    \item \textit{Implementation with strict security proof.} Although our universal criterion provides an idea to detect all attacks covered by our model, whether the verifier will choose the correct honest prover relies on elaborate implementation, such as suitable parameter setting. Considering that the wrong choice in the real world may incur serious legal consequences, to guarantee the correctness of verifier's output, it may require to provide a strict proof of the implementation's security via formal verification.
    \item \textit{Application to real-world large language models (LLMs).}
    Recall that our PoT construction relies on a trajectory matching algorithm to distill synthetic data from the training record. Although our PoT does not limit the range of target AI models, applying it to LLMs such as ChatGPT~\cite{LLMsurvey} may pose additional challenges due to LLMs' unique features. One is that LLMs perform natural language processing (NLP) tasks, whereas most of existing trajectory matching algorithms are designed for computer vision (CV) tasks. Still, recent work~\cite{LLMtrajectorymatching} has identified a candidate for NLP tasks that can help tackle this challenge. The other is that LLMs have an extremely large number of parameters, potentially leading to a high computing workload for verifiers. Existing works explore lightweight trajectory matching algorithms~\cite{LLMlightweight}, which can help to tackle this challenge.
\end{itemize}

\newpage

\bibliographystyle{unsrt}
\bibliography{usenix2024}

\appendix
\section{Trajectory Matching Algorithm}
\label{appendix-code}
\begin{algorithm}[hb]
\caption{Data Synthesis via Trajectory Matching}\label{alg:mtt}
\begin{algorithmic}[1]
\Require $\mathbb{T}_T=\{M_0,\ldots,M_n\}$: prover's length-$n$ trajectory
\Require $\mathbb{T}_{A}$: prover's training algorithms
\Require $\mathcal{D}_I$: distribution of initialized synthetic data
\Require $K$: length of each fragment of trajectory
\Require $T$: \# of iterations for trajectory matching
\Ensure Synthetic Data $S$
\State \text{Initialize synthetic data $S$ by sampling from }$\mathcal{D}_I$
\For{$t=1,\dots,T$}
    \State \text{Sample $i$ so that $0\leq i$ and $i+K\leq n$}
    \State \text{Initialize a synthetic model $M^\prime_0$ as $M_i$}
    \For{$k=0,\dots,K-1$}
    \State Train the synthetic model using $\mathbb{T}_{A}$ and $S$ as \[M^\prime_{k+1}=M^\prime_k+\mathbb{T}_{A,i+k}(M_k^\prime,S)\] 
    \EndFor
    \State Compute the loss function $\mathcal{L}_{\,t}$ between $M^\prime_{K}$ and $M_{i+K}$
    \State Update $S$ with respect to $\mathcal{L}_{\,t}$  
\EndFor

\end{algorithmic}
\end{algorithm}
Our PoT construction relies on a trajectory matching algorithm to generate the synthetic data $S$ from prover's trajectory and training algorithms. 
Although the details of different trajectory matching algorithms may vary, all of these algorithms follow the same paradigm.
Specifically, this paradigm consists of $T$ iterations. In each iteration, the verifier samples a fragment with length $K$ from prover's trajectory, say, from $M_i$ to $M_{i+K}$. Then a synthetic model $M^\prime$ is updated from $M_i$ using the same training algorithms in $\mathbb{T}_A$ but different training data (i.e., using synthetic data $S$ instead of prover's training data). After the ending synthetic model $M_K^\prime$ is generated, the verifier computes some loss function between it and the corresponding model in prover's trajectory $M_{i+K}$. By minimizing such a loss function, the synthetic data is optimized to approach the prover's training data. 
We describe such a paradigm in the form of pseudo-random code in Algorithm \ref{alg:mtt}.

\section{Proof of Theorem 1}
\label{appendix-proof}
\begin{thmprime2}{theo:dis}[Restated]
    Suppose that the malicious prover conducts algorithm manipulation attacks with a lower cost than honest training. Let $(\mathbb{I}_A,\mathbb{T}_A,\mathbb{T}_T,D)$ and $(\mathbb{I}_A,\mathbb{T}_A^{(M)},\mathbb{T}_T^{(M)},D^{(M)})$ denote the training records of the honest prover and the malicious prover, respectively. When Assumptions \ref{assu:strong}-\ref{assu:strong-2} hold, we have    
    \begin{equation}
        I(D;\mathbb{T}_{T,[i:i+k]} )> I(D^{(M)};\mathbb{T}_{T,[i:i+k]}^{(M)}),
    \end{equation}
    where $\mathbb{T}_{T,[i:i+k]}$ and $\mathbb{T}_{T,[i:i+k]}^{(M)}$ are two length-$k$ fragments from $\mathbb{T}_T$ and $\mathbb{T}_{T}^{(M)}$, respectively.
    
    
\end{thmprime2}

\noindent\textit{Proof:} We conduct this proof by 1) computing the mutual information $I(\mathbb{T}_{T,[i:i+k]};D)$ and 2) comparing the values of mutual information between honest prover and manipulated prover.
Before conducting the proof, we provide several basic knowledge on information theory that will be used in the subsequent proof. The information theory has established that for random variables $x,y,z$, the following equations hold~\cite{shannon}:

\begin{align}
    I(x;y)
    =H(x)-H(x\mid y),\label{proof-basic-1}\\
    I(\{x_1,x_2,\ldots,x_n\}; y)=\sum_{i=1}^{n} I(x_i;y\mid \{x_1,x_2,\ldots,x_{i-1}\})\label{proof-basic-chain}.
\end{align}
Particularly, if $z$ is independent with $x$ conditioning on $y$ (i.e., $I(x,z\mid y)=0$), then we have:
\begin{equation}
    H(x\mid y,z)=H(x\mid y).\label{proof-basic-3}
\end{equation}

\noindent\textbf{1) Mutual Information Computation.} By Equations \eqref{proof-basic-1} and \eqref{proof-basic-chain}, we have the following equations for training trajectory $\mathbb{T}_T$ and datasets $D$.
\begin{align}
    &I(\mathbb{T}_{T,[i:i+k]};D)\\
    =&I(\{M_i,\ldots,M_{i+k}\};D)\\
    =&\sum_{j=i}^{i+k-1} I(M_j;D\mid \{M_i,\ldots,M_{j-1}\})\\
    =&\sum_{j=i}^{i+k-1} H(M_i\mid \{M_1,\ldots,M_{j-1}\})-H(M_i\mid \{D,M_1,\ldots,M_{j-1}\}).\label{proof-basic-4}
\end{align}

In the case where $\mathbb{T}_T$ is generated by structurally correct training algorithms, the $j$-th model state $M_j$ is trained from the $(j-1)$-th state $M_{j-1}$, but does not use previous states $M_1,\ldots,M_{j-1}$.
Hence, conditioning on the knowledge of $M_{j-1}$, model state $M_j$ is independent from $\{M_1,\ldots,M_{j-2}\}$. By Equation \eqref{proof-basic-3}, we have for $1\leq j\leq n$,
\begin{equation}
    H(M_j\mid \{M_i,\ldots,M_{j-1}\}) = H(M_j\mid M_{j-1}).\label{proof-case1-step1-1.1}
\end{equation}
By conditioning on $D$, we have the following equation from Equation \eqref{proof-case1-step1-1.1}.
\begin{equation}
    H(M_j\mid \{D,M_i,\ldots,M_{j-1}\}) = H(M_j\mid \{D,M_{j-1}\}).\label{proof-case1-step1-1.2}
\end{equation}
By substituting Equations \eqref{proof-case1-step1-1.1} and \eqref{proof-case1-step1-1.2} into Equation \eqref{proof-basic-4}, we have
\begin{align}
    &I(\mathbb{T}_{T,[i:i+k]};D)\\
    =&\sum_{j=i}^{i+k-1} H(M_i\mid \{M_1,\ldots,M_{j-1}\})-H(M_i\mid \{D,M_1,\ldots,M_{j-1}\})\\
    =&\sum_{j=i}^{i+k-1} H(M_j\mid M_{j-1})-H(M_j\mid \{D,M_{j-1}\}).\label{proof-mi-honest-1}
\end{align}

For honest prover, the training data can honestly generate the trajectory through training algorithms, i.e., $M_{i+1}=\mathbb{T}_{A,i}(M_{i},D_i)+z_i$. By substituting it into Equations \eqref{proof-mi-honest-1}, we have

\begin{align}
    &H(M_j\mid M_{j-1})\\
    =&H(\mathbb{T}_{A,j-1}(M_{j-1},D_{j-1})+z_{j-1}\mid M_{j-1})\\
    =&H(\mathbb{T}_{A,j-1}(M_{j-1},D_{j-1})\mid M_{j-1})+H(z_{j-1}),\label{proof-honest-stage3}
\end{align}
where Equation \eqref{proof-honest-stage3} is due to the independence between $z_{j-1}$ and $D_{j-1},M_{j-1}$.
By conditioning on $D$, we have the following equation from Equation \eqref{proof-honest-stage3}.
\begin{align}
    &H(M_j\mid \{D,M_{j-1}\})\\
    =&H(\mathbb{T}_{A,j-1}(M_{j-1},D_{j-1})\mid \{D,M_{j-1}\})+H(z_{j-1}\mid D).\label{proof-honest-stage4}
\end{align}
Note that conditioning on the knowledge of $\{D,M_{j-1}\}$, the output of $\mathbb{T}_{A,j}(M_{j-1},D_j)$ is a deterministic value with no uncertainty. Hence, we have $H(\mathbb{T}_{A,j-1}(M_{j-1},D_{j-1})\mid \{D,M_{j-1}\})=0$. In addition, since $z_{j-1}$ is a random noise independent from $D$, we have $H(z_j\mid D)=H(z_j)$ by Equation \eqref{proof-basic-3}. By substituting these two conclusions into \eqref{proof-honest-stage4}, the following equation holds:
\begin{align}
    H(M_j\mid \{D,M_{j-1}\})=H(z_{j-1}).\label{proof-honest-stage5}
\end{align}

By substituting Equations \eqref{proof-honest-stage3} and \eqref{proof-honest-stage5} into Equation \eqref{proof-mi-honest-1}, we have
\begin{align}
I(\mathbb{T}_{T,[i:i+k]};D)
    =&\sum_{j=i}^{i+k-1} H(\mathbb{T}_{A,j-1}(M_{j-1},D_{j-1})\mid M_{j-1})\label{proof-mi-honest-part}
\end{align}

Note that the malicious prover conducts algorithm manipulation attacks, by which the training data are also honest. Therefore, the above derivation also applies to the malicious prover. Similarly, we have
\begin{align}
I(\mathbb{T}^{(M)}_{T,[i:i+k]};D^{(M)})
    =&\sum_{j=i}^{i+k-1} H(\mathbb{T}^{(M)}_{A,j-1}(M^\prime_{j-1},D^{(M)}_{j-1})\mid M^\prime_{j-1})\label{proof-mi-malicious-part}
\end{align}


\noindent\textbf{2) Mutual Information Comparison.} Next, by comparing the mutual informations between honest training records and forged training records, we come to the conclusion of Theorem 1, i.e., $I(D;\mathbb{T}_{T,[i:i+k]} )> I(D^{(M)};\mathbb{T}_{T,[i:i+k]}^{(M)})$.
Recall that the algorithm manipulation attacks can be conducted along forward- and reverse- two directions. 
We prove this conclusion along each direction, respectively.

\noindent\textbf{Forward-Direction.}
When the cost of malicious prover is less than that of the honest prover, by Assumption $1$, the forward algorithm manipulation attacks must utilize the target model $M$ in the training algorithms $\mathbb{T}^{(M)}_{A,j-1}$. Hence, compared with honest training algorithms, the output of manipulated training algorithms has lower entropy that that of honest prover, i.e., 

\begin{align}
    H(\mathbb{T}_{A,j-1}&(M_{j-1},D_{j-1})\mid M_{j-1})> \\
    & H(\mathbb{T}^{(M)}_{A,j-1}(M^\prime_{j-1},D^{(M)}_{j-1})\mid M^\prime_{j-1}).\label{proof-forward-1}
\end{align}

\noindent\textbf{Reverse-Direction.}
By our modeling of reverse algorithm manipulation attacks, when the cost of malicious prover is lower than that of honest prover, the output of $\mathbb{T}_{A,i}^{(M)}$ is a deterministic value and thus has entropy being close to zero. In contrast, the entropy of $\mathbb{T}_{A,i}^{(M)}$'s output is far greater than zero. Hence, along the reverse direction, we also has
\begin{align}
    H(\mathbb{T}_{A,j-1}&(M_{j-1},D_{j-1})\mid M_{j-1}) \\
    >0 \approx & H(\mathbb{T}^{(M)}_{A,j-1}(M^\prime_{j-1},D^{(M)}_{j-1})\mid M^\prime_{j-1}).\label{proof-forward-2}
\end{align}

By combining Equations \eqref{proof-mi-honest-part} -- \eqref{proof-forward-2}, we come to the conclusion $I(M_j^\prime;D^{(M)}\mid M_{j-1}^\prime)\leq I(M_i;D\mid M_{i-1})$ along the forward- and reverse- directions. The proof is complete.
$\hfill\blacksquare$

\end{document}